\documentclass[twocolumn,trackchanges]{aastex63}

\newcommand{\mytilde}{\raise.19ex\hbox{$\scriptstyle\sim$}}
\turnoffedit
\submitjournal{ApJ}

\shorttitle{Merging Galaxy Cluster CIZAJ0107}
\shortauthors{Finner et al.}
\graphicspath{{./}{figures/}}
\begin{document}

\title{HST and HSC Weak-lensing Study of the Equal-mass Dissociative Merger CIZA J0107.7+5408}

\correspondingauthor{Kyle Finner}
\email{kfinner@caltech.edu}

\author[0000-0002-4462-0709]{Kyle Finner}
\affiliation{Harvard-Smithsonian Center for Astrophysics, 60 Garden St., Cambridge, MA 02138, USA}
\affiliation{Infrared Processing and Analysis Center, California Institute of Technology, 1200 E California Blvd., Pasadena, CA 91125, USA}
\affiliation{Department of Astronomy, Yonsei University, 50 Yonsei-ro, Seodaemun-gu, Seoul 03722, Republic of Korea}

\author[0000-0002-3984-4337]{Scott W. Randall}
\affiliation{Harvard-Smithsonian Center for Astrophysics, 60 Garden St., Cambridge, MA 02138, USA}

\author[0000-0002-5751-3697]{M. James Jee}
\affiliation{Department of Astronomy, Yonsei University, 50 Yonsei-ro, Seodaemun-gu, Seoul 03722, Republic of Korea}
\affiliation{Department of Physics and Astronomy, University of California, Davis, One Shields Avenue, Davis, CA 95616, USA}

\author[0000-0002-0485-6573]{Elizabeth L. Blanton}
\affiliation{Institute for Astrophysical Research and Astronomy Department, Boston University, 725 Commonwealth Avenue, Boston, MA 02215, USA}

\author[0000-0001-5966-5072]{Hyejeon Cho}
\affiliation{Department of Astronomy, Yonsei University, 50 Yonsei-ro, Seodaemun-gu, Seoul 03722, Republic of Korea}
\affiliation{Center for Galaxy Evolution Research, Yonsei University, 50 Yonsei-ro, Seodaemun-gu, Seoul 03722, Republic of Korea}

\author[0000-0001-6812-7938]{Tracy E. Clarke}
\affiliation{Naval Research Laboratory, Code 7213, 4555 Overlook Ave. SW, Washington, DC 20375, USA}

\author[0000-0002-1634-9886]{Simona Giacintucci}
\affiliation{Naval Research Laboratory, Code 7213, 4555 Overlook Ave. SW, Washington, DC 20375, USA}

\author[0000-0003-0297-4493]{Paul Nulsen}
\affiliation{Harvard-Smithsonian Center for Astrophysics, 60 Garden St., Cambridge, MA 02138, USA}
\affiliation{ICRAR, University of Western Australia, 35 Stirling Hwy, Crawley, WA 6009, Australia}

\author[0000-0002-0587-1660]{Reinout van Weeren}
\affiliation{Leiden Observatory, Leiden University, P.O. Box 9513, NL-2300 RA Leiden, the Netherlands}

\begin{abstract}
A dissociative merger is formed by the interplay of ram pressure and gravitational forces, which can lead to a spatial displacement of the dark matter and baryonic components of the recently collided subclusters. CIZA J0107.7+5408 is a nearby (z=0.105) dissociative merger that hosts two X-ray brightness peaks and a bimodal galaxy distribution. Analyzing MMT/Hectospec observations, we investigate the line-of-sight and spatial distribution of cluster galaxies.
Utilizing deep, high-resolution \textit{Hubble} Space Telescope Advanced Camera for Surveys imaging and large field-of-view Subaru Hyper-Suprime-Cam observations, we perform a weak-lensing analysis of CIZA J0107.7+5408. Our weak-lensing analysis detects a bimodal mass distribution that is spatially consistent with the cluster galaxies but significantly offset from the X-ray brightness peaks. Fitting two NFW halos to the lensing signal, we find an equal-mass merger with subcluster masses of \edit2{$M_{200,NE}=2.8^{+1.1}_{-1.1}\times10^{14}$ M$_\odot$} and \edit2{$M_{200,SW}=3.1^{+1.2}_{-1.2}\times10^{14}$ M$_\odot$}. Moreover, the mass-to-light ratios of the subclusters, \edit2{$(M/L)_{NE}=571^{+89}_{-91}\ M_\sun/L_{\sun,B}$} and \edit2{$(M/L)_{SW}=564^{+87}_{-89}\ M_\sun/L_{\sun,B}$}, are found to be consistent with each other and within the range of mass-to-light ratios found for galaxy clusters.

\end{abstract}



\section{Introduction} \label{sec:intro}

A class of merging galaxy clusters, \textit{dissociative mergers} \citep{2012dawson}, are ideal laboratories to investigate the nature of dark matter. Dissociative mergers exhibit significant spatial separation of their baryon and dark matter components. The separations are found in post-pericenter cluster-cluster mergers and are caused by the differing collisional properties of dark matter and the intracluster medium (ICM). The degree of gas dissociation depends on the density of the colliding subclusters \citep{2000burkert}. Therefore, the greatest gas dissociation occurs for head-on collisions. The offset of the gas from the dark matter halo reaches a maximum during the period between first pericenter and first apocenter \citep{2017kim}. Since the ICM is gravitationally bound to the hosting dark matter halo, the ICM-dark matter offset eventually disappears. Thus, gas-dark matter offsets have a short lifetime. Furthermore, dissociative mergers are best viewed in plane-of-the-sky mergers, which limit projection effects. The geometric constraints and the short lifetime of the baryon-dark matter offsets make observations of dissociative mergers rare.

The most renowned dissociative merger is the Bullet cluster at $z=0.296$ \citep{2004markevitch}. The large offset of the weak-lensing peak and X-ray emitting ICM of the Bullet cluster provided strong evidence for the existence of dark matter \citep{2006clowe}. The dissociated configuration of gas from the dark matter and galaxies allowed constraints to be placed on the self-interacting cross-section of dark matter \citep{2004markevitch, 2008randall}. Additional dissociative mergers have been found and used to constrain the self-interacting cross-section of dark matter \cite[e.g.][]{2008bradac, 2012dawson}. Since dissociative mergers are rare, increasing the sample size of clean mergers is of the utmost importance.

\cite{2016randall} announced the discovery of a nearby ($z=0.105$) dissociative merger, CIZA J0107.7+5408 (CIZAJ0107 hereafter). Their X-ray (\textit{Chandra}), radio (VLA\footnote{Very Large Array}, WSRT\footnote{Westerbork Synthesis Radio Telescope}, GMRT\footnote{Giant Metrewave Radio Telescope}), and optical (INT\footnote{Isaac Newton Telescope}) analysis of CIZAJ0107 characterized the merging system. They observed X-ray emission with an elongated morphology in the northeast-southwest direction and two brightness peaks. The study highlighted the prominent separation of the X-ray brightness peaks from the brightest cluster galaxies (BCGs) and from the cluster galaxy light distribution. They measured the global temperature of the cluster to be $kT=7.8^{+0.5}_{-0.5}$~keV and estimated the total mass to be $M_{500}=7.8^{+0.8}_{-0.7}\times10^{14}\ M_\odot$ by the $M_{500}-T_X$ scaling relation. This mass estimate is higher than the \textit{Planck} Sunyaev-Zel'dovich estimate of $M_{500,SZ}=5.8^{+0.3}_{-0.3}\times10^{14}\ M_\odot$ \citep{2014planck}.

\cite{2016randall} also pointed out features that indicate a post-merger system. A high-temperature and -pressure region was detected in the southwest of the cluster and in line with the merger axis defined by the two subcluster BCGs. In this region, the temperature jumps to $18.5 \pm 6.0$~keV, which they postulate is a merger-induced shock with a Mach number $M=2.3\pm0.4$. Their analysis of 1.4 GHz WSRT radio observations detected diffuse radio emission in the northern and southern regions of the cluster. The location of the southern radio emission coincides with the hot region of the ICM, making it a candidate radio relic. Radio relics are synchrotron emission from cosmic rays (re-)accelerated in merger shocks \citep[see][for a review]{2019vanweeren}.

CIZAJ0107 has key differences from the Bullet cluster. The two X-ray brightness peaks of CIZAJ0107 are of similar brightness, whereas, in the Bullet cluster, the Bullet is significantly brighter. Therefore, it may be expected that CIZAJ0107 is a more equal-mass merger than the Bullet cluster, which is believed to be a \mytilde10:1 merger. Other differences are that CIZAJ0107 does not host a cool core and that the viewing angle of the merger may depart from the plane of the sky. Some of the evidence that suggests CIZAJ0107 is not a plane-of-the-sky merger are the differing BCG redshifts and the complex morphology of the candidate radio relic \citep{2016randall}. These differences will be explored in our analysis.

The dissociative nature of CIZAJ0107 and its distinct merger features signify that CIZAJ0107 is an important cluster for probing dark matter. Characterizing the mass of the subclusters is critical for reconstructing the merger scenario and to study the properties of dark matter. However, an understanding of the underlying mass distribution is still lacking. The goals of this study are to use weak lensing to map the mass distribution, identify mass substructures, and estimate their masses. As a step towards constraining dark matter, we will quantify the locations of the weak lensing and galaxy peaks, as well as compare the mass-to-light ratios of the subclusters. These goals are dependent on deep and wide observations to enable a high-fidelity weak-lensing analysis over a large field of view. The combination of the high-resolution HST observations with wide field-of-view Subaru Hyper-Suprime-Cam (HSC) observations is essential for constraining the mass peaks while also constraining the total mass of the cluster without extrapolation.

In this work, we use the common notation where the symbol $R_{200}$ denotes the radius of the sphere within which the average density is 200 times the critical density of the universe. A flat cosmology is assumed with $H_0=70$ km s$^{-1}$ Mpc$^{-1}$, $\Omega_m=0.3$, and $\Omega_\Lambda=0.7$. At the redshift of CIZAJ0107 ($z=0.105$), the physical to angular conversion is 1.93 kpc/\arcsec. All magnitudes are reported in the AB magnitude system and all uncertainties are one sigma (68\% probability), unless otherwise noted.

\begin{figure*}[!ht]
    \centering
    \includegraphics[width=\textwidth]{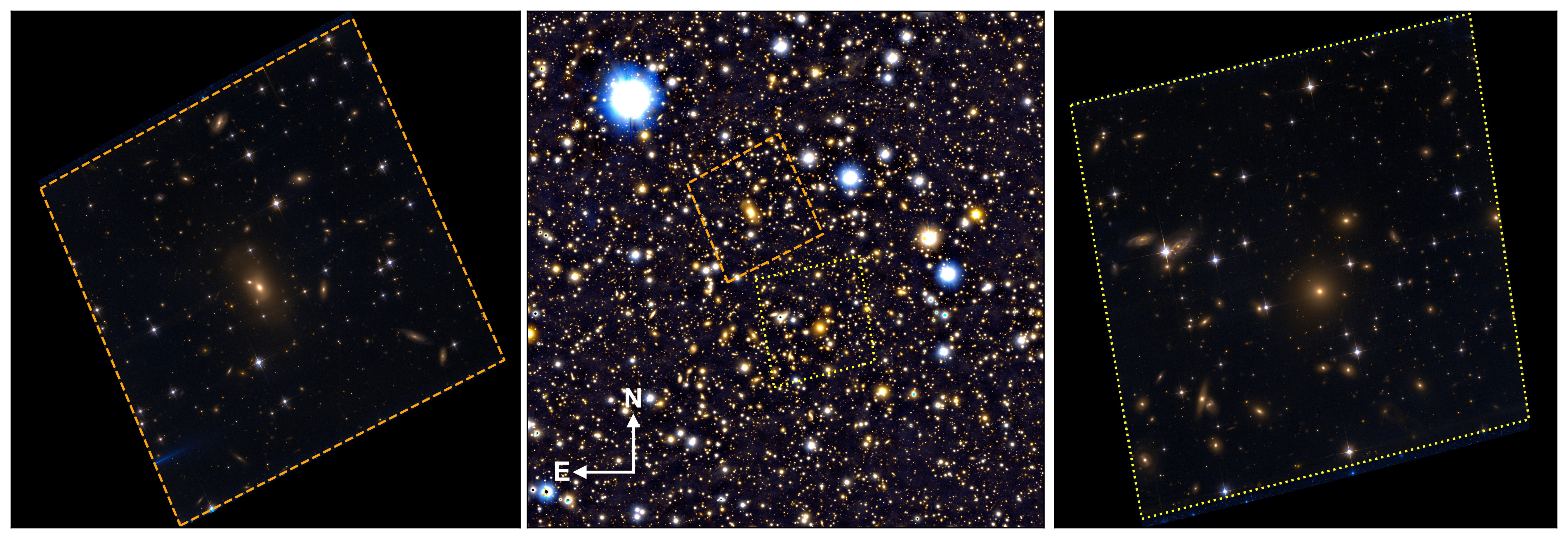}
    \caption{Color images of CIZAJ0107. Left: HST pointing of NE subcluster. Middle: HSC view of the cluster. Orange and yellow HST footprints match the left and right panels. Right: HST pointing of the SW subcluster. Filters for HST color images are r=F814W, g=F606W+F814W, and b=F606W and for the HSC color image are r=\textit{HSC-r2}, g=(\textit{HSC-g})+(\textit{HSC-r2}), and b=\textit{HSC-g}.}
    \label{fig:color_image}
\end{figure*}
\section{Observations and Data Reduction} \label{sec:data_red}
\subsection{\textit{Hubble} Space Telescope}\label{sec:hst_obs}
HST observations were completed on 2019 March 15 and 2019 October 18 under GO program 15610 (PI: S. Randall). The northeast and southwest cluster regions, shown in Figure \ref{fig:color_image}, were each observed with the Advanced Camera for Surveys (ACS) for total exposures of 4800s in F814W and 2352s in F606W. 

\edit2{The calibrated FLC files were downloaded from the Mikulski Archive for Space Telescopes and are available at\dataset[doi:10.17909/d3ny-rj52]{https://doi.org/10.17909/d3ny-rj52}}. The astrometric solution of the FLC files was refined by finding common objects within frames, deriving the optimal alignment, and applying the shift to the world coordinate system (WCS). The well-aligned frames were then coadded into a mosaic with a final pixel scale of 0\farcs05 per pixel. The coadd was achieved with Astrodrizzle \citep{2012gonzaga} utilizing a Lanczos3 kernel. This reduction method for ACS imaging has been shown to be optimal for lensing studies \citep{2014jee}. 

\subsection{Subaru}\label{sec:hsc_obs}
The Subaru Hyper Suprime-Cam (HSC) was used to observe CIZAJ0107 on 2020 November 13, 2020 November 14, and 2020 November 20 (PI: K. Finner). Total exposure times of 6840s in \textit{HSC-r2} and 1440s in \textit{HSC-g} were obtained. The observation plan utilized single exposures of 180s integration with a rotation between each exposure to lessen the effect of nearby bright stars. Rotating the camera between exposures is advantageous to weak-lensing analysis because it diminishes bleeding trails and diffraction spikes.

Single-frame processing (bias, flat, dark, etc.) was done with the HSCPipe \citep{2018bosch}. The calibrated fits files are output with a WCS following the simple imaging polynomial (SIP) format for geometric distortion. To be compatible with SCAMP, the headers were transformed from SIP to TPV with the publicly available sip\_tpv code\footnote{https://github.com/stargaser/sip\_tpv} \citep{2012shupe}. SCAMP \citep{bertinscamp} was then used to align the astrometry of the processed frames to the Pan-STARRS-DR1 catalog \citep{2020panstarrs}. The single frames were stacked into a median-coadded mosaic with SWarp \citep{bertinswarp}. Using the median image as a guide, $3\sigma$ outlying pixels were flagged in weight files which were then used to create a mean-coadded mosaic by a second SWarp run.

For both the HST and HSC imaging, object detection was performed with SExtractor \citep{1996bertin} in dual-image mode. We chose the deeper image (\textit{HSC-r2}, F814W) from each telescope as the detection image. For HST, the photometric zeropoints were calibrated from the value retrieved from the ACS Zeropoints Calculator\footnote{https://acszeropoints.stsci.edu/}. The \textit{HSC-g} and \textit{HSC-r2} zeropoints were calibrated from the magnitude of matched stars in the Pan-STARRS-DR1 catalog $g$- and $r$-band photometry. In this work, colors are derived from the difference of MAG\_ISO magnitudes and total magnitudes are from MAG\_AUTO. 

\subsection{MMT}\label{sec:mmt}
Hectospec fiber spectroscopic observations of cluster-galaxy candidates were gathered from the 6.5m MMT on Mount Hopkins. Two configurations of the 300 fiber spectrograph were designed to sample the galaxy population of CIZAJ0107. The observations were successfully completed on 2015 September 13 (PI: S. Randall).

The spectra were reduced with the HSRED2.0 software\footnote{http://www.mmto.org/hsred-reduction-pipeline/} that is provided by the MMT. The reduced spectra are sky subtracted, variance weighted, and coadded to a science-ready product. Redshifts were determined by cross-correlating the observed spectra with template spectra using RVSAO \citep{1998kurtz}. Following \cite{1998kurtz}, we only keep redshifts that the RVSAO script deems are secure (r-value above 4). 

\section{Spectroscopic Analysis}\label{sec:spec_analysis}
The wide field of view ($1^\circ$ diameter) of the Hectospec instrument is complementary to the field of view of Subaru HSC. From the two pointings of the MMT, we recovered \edit2{124} secure redshifts with r-value $>4$. 

\begin{figure}
    \centering
    \includegraphics[width=0.48\textwidth]{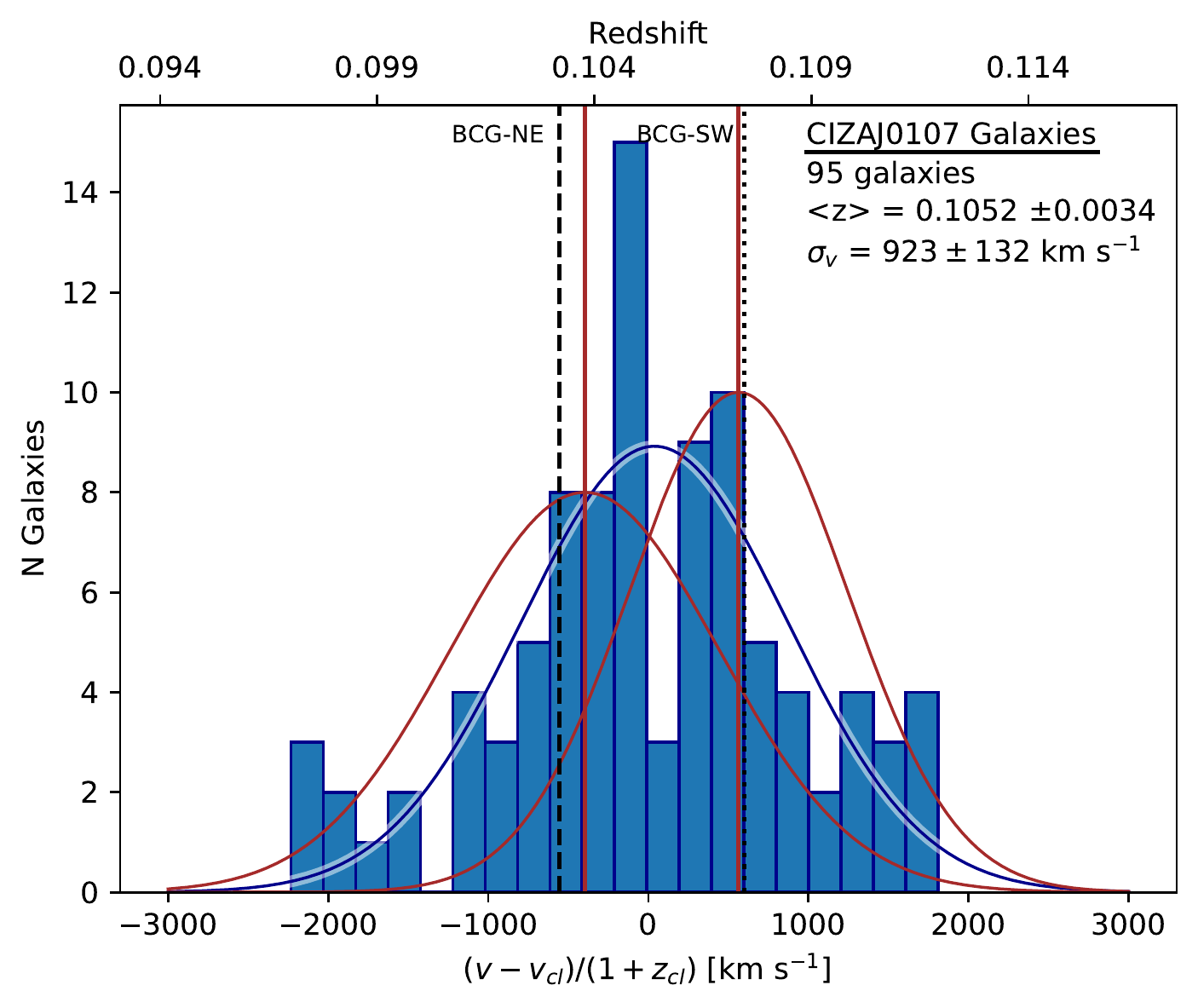}
    \caption{Relative radial velocity histogram of the \edit2{95} galaxies within $3\sigma$ of the average cluster-galaxy redshift. The best-fit Gaussian is shown as a blue curve. The redshift of the northern (southern) BCG is marked with a black-dashed (-dotted) vertical line. A two-component Gaussian fit is colored brown with vertical brown lines indicating the mean of each distribution. The two components have mean values of \edit2{$-412$ km s$^{-1}$} and  \edit2{$546$ km s$^{-1}$} with standard deviations of  \edit2{$839\pm124$ km s$^{-1}$} (with mean closest to NE BCG) and \edit2{$681\pm106$ km s$^{-1}$} (with mean closest to SW BCG).   The two-component Gaussian fit is disfavored by BIC score.}
    \label{fig:spec_members}
\end{figure}

To select cluster member galaxies, we first gathered all galaxies with radial velocity within a $\pm4000$ km s$^{-1}$ window centered on $z=0.107$ \citep{2002ebeling} and projected distance within 3 Mpc of the midpoint between the cluster BCGs. The velocity and projected distance windows were chosen to be larger than the expected $3\sigma_v$ (velocity dispersion) limit and the expected $R_{200}$ based on the mass estimated in \cite{2016randall}. From this selection, we refined the average redshift to be \edit2{$z=0.1052\pm0.0034$} and re-selected galaxies with the window centered on the average redshift. This method resulted in a final selection of \edit2{95} galaxies, which are presented in Figure \ref{fig:spec_members}. The BCG redshifts are marked in the figure with dashed (NE; $z=0.1034\pm0.00007$) and dotted (SW; $z=0.1076\pm0.00006$) vertical black lines. The velocity difference of the BCGs is \mytilde1250 km s$^{-1}$. Utilizing the bi-weight estimator \citep{1990beers}, we determined the line-of-sight velocity dispersion of the cluster as a whole to be \edit2{$\sigma_v=923\pm132$ km s$^{-1}$}.

As an attempt to separate the galaxies into two clusters, the relative radial velocities of the galaxies were fit with a two-component Gaussian distribution (brown curves in Figure \ref{fig:spec_members}). In Figure \ref{fig:spec_members}, the mean of each distribution is marked by a brown vertical line. Both means are within 100 km s$^{-1}$ of the nearest BCG radial velocity, with \edit2{-412 km s$^{-1}$} close to the NE BCG and \edit2{546 km s$^{-1}$} close to the SW BCG. The standard deviations of the Gaussian distributions are \edit2{$839\pm124$ km s$^{-1}$} (with mean closest to NE BCG) and \edit2{$681\pm106$ km s$^{-1}$} (with mean closest to SW BCG). Following the technique applied in \cite{2019bgolovich}, we calculate the Bayesian information criterion (BIC) scores and use them as a test for the best-fit model. The BIC score penalizes models with more parameters to prevent over fitting. The two-component Gaussian fit is disfavored  \edit2{($BIC_1-BIC_2=14$)} compared to the one-component fit (blue curve). Using the relative Akaike information criteria, we find that the one-component model is \edit2{$EXP(6.0/2)\simeq20$} times more likely than the two-component model. Attempting a three-dimension (R.A.,Dec.,velocity), two-component, Gaussian-mixture-model fit failed to separate the galaxies into two meaningful clusters.

\section{Weak-lensing Method}\label{sec:wl_method}
\subsection{Basic Weak-lensing Theory}\label{sec_wl_theory}
In weak-gravitational-lensing, intervening mass, a galaxy cluster in this case, deflects the light from background galaxies, causing minute distortions in their images. Detecting the weak-lensing effect requires a statistical analysis of the observed shapes of background galaxies. The linearized distortion of the galaxy image follows the transformation

\begin{equation}
    A = (1-\kappa)\left( \begin{array}{cc}
 1 - g_1 & -g_2 \\
 -g_2 & 1 + g_1 \\
 \end{array}  \right). \label{eqn_A}
\end{equation}
The convergence $\kappa=\Sigma/\Sigma_c$ is the projected mass density $\Sigma$ normalized by the critical surface density $\Sigma_c$:

\begin{equation}\label{eq:critical_density}
    \Sigma_c = \frac{c^2}{4\pi G D_l}\frac{D_s}{D_{ls}},
\end{equation}
where $c$ is the speed of light, $G$ is the gravitational constant, and $D_l$, $D_s$, and $D_{ls}$ are the angular diameter distances from observer to gravitational lens, observer to source galaxy, and gravitational lens to source galaxy.

The reduced shear $g=g_1+ig_2$ is an anisotropic distortion that transforms circles into ellipses. The reduced shear can be measured by determining the average ellipticity of galaxy images. We define the ellipticity as $e=(a-b)/(a+b)$, where $a$ is the semi-major axis of the ellipse and $b$ is the semi-minor axis. The ellipticity and orientation of a galaxy image can be represented in the complex number $\epsilon=e_1+ie_2$ with $e_1=e\cos{2\theta}$ and $e_2=e\sin{2\theta}$, where $\theta$ is the position angle of the major axis.  

\subsection{PSF Modeling and Shape Fitting}\label{sec:psf_modeling}
PSF modeling for the HST and HSC observations was achieved by a Principal Component Analysis (PCA) of the observed stars. The technique was first described for HST ACS imaging in \cite{2007jee}. Beyond the HST ACS, the technique has been applied to observations with HST WFC3 \citep{2020finner}, Blanco DECam \citep{2020hyeonghan}, Subaru Suprime-Cam \citep{2015jee, 2017finner, 2021finner, 2022Cho}, and now Subaru HSC (this work). We refer the reader to the aforementioned literature for details on the method.

The weak-lensing technique applied in this work forward models a PSF while model fitting each galaxy on the coadded mosaics. The coadded mosaics inherit the PSF of the individual frames that comprise them. Therefore, we model the PSF for each object in each frame and then stack the PSFs for use with the coadded mosaic. This method is required because the orientation and size of the PSF across the telescope field of view is complex and it is discontinuous from CCD to CCD \citep{2011jeeLSST}.

We use an elliptical Gaussian function to model the light profile of each galaxy. The seven parameters of the elliptical Gaussian function are background, amplitude, position x and y, variance $\sigma_x^2$ and $\sigma_y^2$, and position angle $\theta$. To improve the fit, we fix the background and position of the model to the values found in SExtractor, leaving four free parameters. Fitting is achieved with the MPFIT code \citep{2009markwardt}, which returns a flag of 1 for well-fit objects. Our image simulations have shown that the elliptical Gaussian function underestimates the ellipticity of galaxies. We apply a multiplicative bias calibration to the measured ellipticities of 1.15 \citep{2016jee} and 1.11 \citep{2014jee} for HSC and HST, respectively. These shear calibration factors were derived following the SFIT technique outlined in \cite{2013jee}. \edit2{We ran additional shear calibration simulations with SFIT and verified that the HSC calibrations are consistent with the Suprime-Cam for our weak-lensing pipeline.}

\begin{figure}[!ht]
    \centering
    \includegraphics[width=0.45\textwidth]{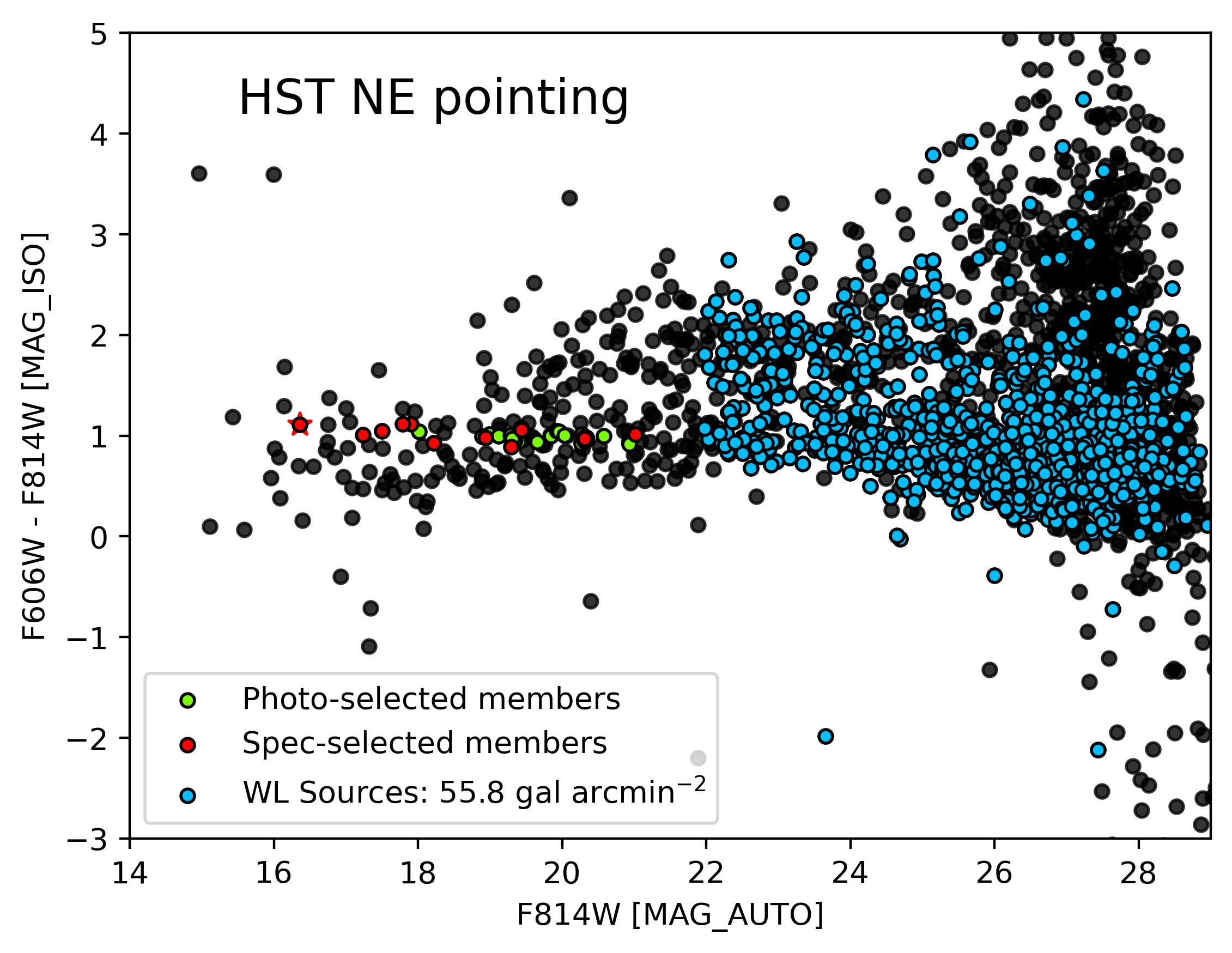}\\
    \includegraphics[width=0.45\textwidth]{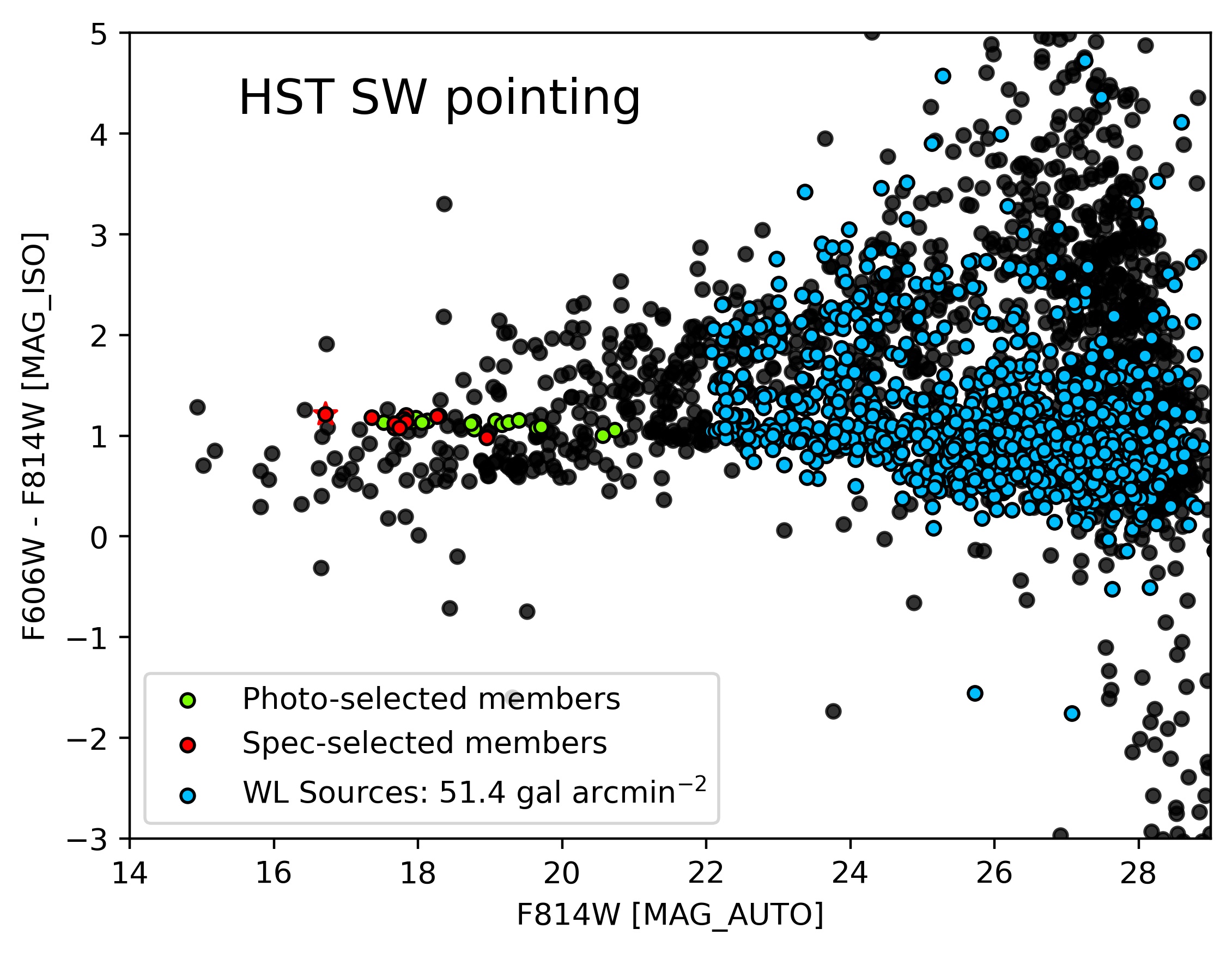}\\
     \includegraphics[width=0.45\textwidth]{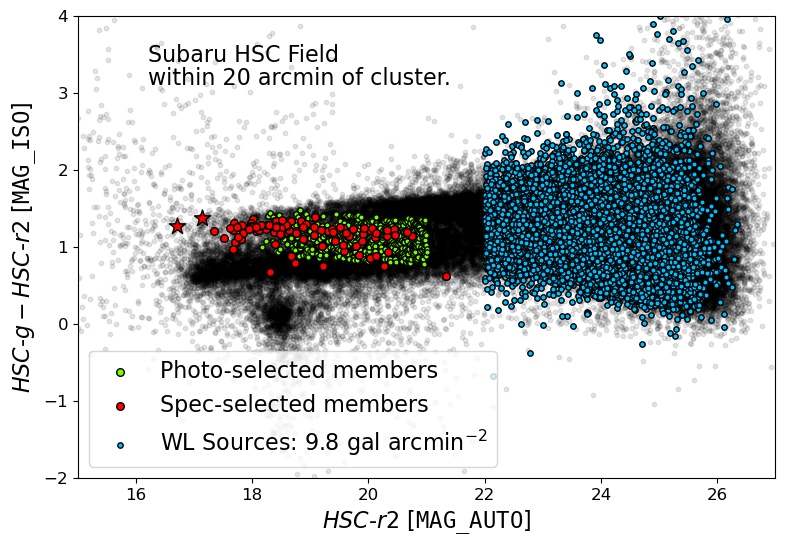}
    \caption{Color-magnitude diagrams for HST NE (top), HST SW (middle), and Subaru HSC observations (bottom). The red stars represents the BCGs. The red circles are spectroscopically confirmed cluster galaxies that were discussed in Section \ref{sec:spec_analysis}. The green circles are candidate cluster galaxies, selected for proximity to the cluster red sequence defined by a linear fit. The blue circles are weak-lensing source galaxies (selection described in Section \ref{sec:source_selection}).}
    \label{fig:cmds}
\end{figure}

\subsection{Source Selection}\label{sec:source_selection}
The weak-lensing signal is only present in the light of the galaxies that are behind the galaxy cluster. Our goal for source selection was to create a source catalog that contains as many background galaxies as possible while limiting the contamination of foreground and cluster galaxies. The field of CIZAJ0107 does not have significant spectroscopic- or photometric-redshift measurements to determine distances, which would simplify background galaxy selection. Instead, we relied on the photometry and shapes of objects to identify background galaxies. We created separate source catalogs for the HST and HSC imaging and then merged them by matching sources and allowing the HST objects to take precedence over the HSC.

Since CIZAJ0107 is a low-redshift cluster, the majority of the galaxies are background to the cluster. Figure \ref{fig:cmds} presents the color-magnitude diagrams (CMDs) for the HST NE (top) and SW (middle) fields and the larger field of view of the HSC (bottom). The spectroscopically selected galaxies from Section \ref{sec:spec_analysis} are marked with red circles and define the red sequence. In some cases, the tight relation of the red sequence can be used as a guide to avoid contaminating the source catalog with cluster galaxies. For example, in high-redshift clusters, the red sequence is used to select the bluer background galaxies \citep[][]{2011jee, 2018schrabback}. However, at low redshift, a significant fraction of the background galaxies are redder than the red sequence. For the lower redshift clusters of LoCuSS ($0.15 < z < 0.3$), \cite{2010okabe} include red and blue galaxies in their source catalog. As CIZAJ0107 is a lower redshift cluster, we decided to include red and blue galaxies in our source catalog. 

To stay clear of the spectroscopically confirmed cluster members, we selected background galaxies by constraining their magnitudes to be \textit{HSC-r2} $>$ 22 for HSC and F814W $>$ 22 for HST. Testing these cuts on the GOODS-S and GOODS-N photometric redshift catalogs, we expect the contamination by foreground galaxies to be below $2\%$. Further refinement of the catalog was achieved through the shape properties of the objects. We rejected galaxies with $e>0.9$ because their high ellipticities are mostly caused by a failure in shape measurement. An ellipticity uncertainty cut of $\sigma_e < 0.3$ was made to remove objects that have a poor elliptical-Gaussian fit. The ellipticity uncertainty criterion is effective at removing stars because their pre-PSF shape is a delta function, which is poorly fit with a Gaussian distribution. Since CIZAJ0107 is located behind a crowded star field, we took extra precaution and limited the FLUX\_RADIUS of sources to greater than 1.5 pixels for the HST and 3.4 pixels for the HSC. This is \mytilde0.2 pixels larger than the size of stars (the PSF) and lessens the chance of unidentified stars contaminating the source catalog. 

\begin{figure}
    \centering
    \includegraphics[width=0.45\textwidth]{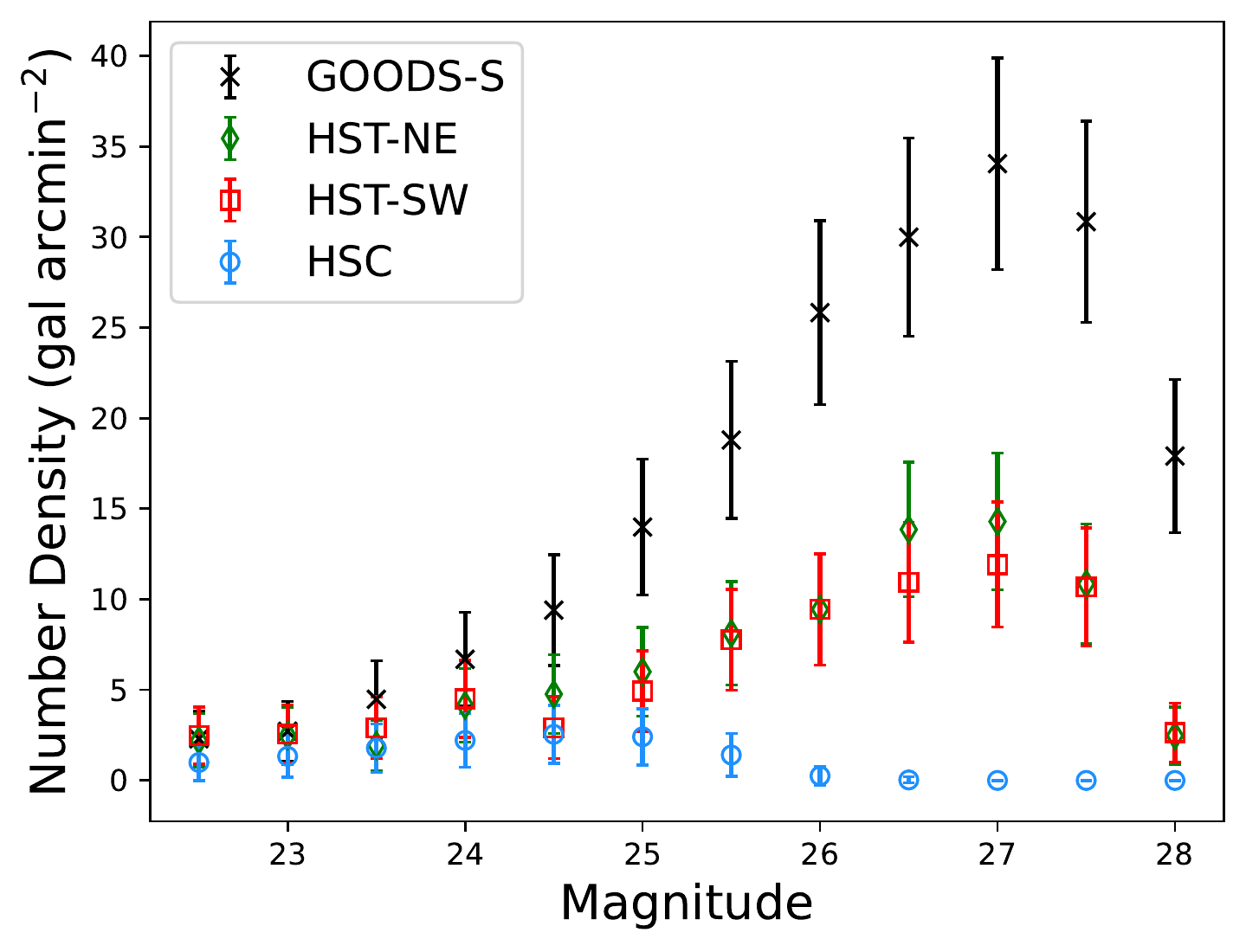}
    \caption{Galaxy number density in magnitude bins for the GOODS-S field and for our CIZAJ0107 source catalogs. At the bright end (Mag $<24$), the four catalogs are consistent. At fainter magnitudes, a fair comparison cannot be made because the GOODS-S is much deeper imaging. The lower number density for the HSC imaging is likely caused by the crowded star field. The ratios of magnitude bins are used in Section \ref{sec:redshift_determination} to estimate the effective redshift and $\beta$ values.}
    \label{fig:gal_hist}
\end{figure}

\edit1{Contamination of the source catalog by foreground and cluster galaxies can lower the weak-lensing signal. We performed three tests for contamination in the source catalog.
1) We examined the number density of source galaxies in radial bins centered at the BCG of each cluster. The expectation is that a significant number of cluster galaxies in the source catalog will lead to an overdensity near the cluster center. A competing effect is from magnification bias, which lowers the number of background galaxies detected. However, the level of magnification bias expected for CIZAJ0107 is relatively low because the cluster is low redshift and is not extremely massive. Our test returned a statistically flat number density relation, which indicates minimal contamination by cluster galaxies. 2) We compared the number density of galaxies as a function of magnitude to the GOODS-S catalog (Figure \ref{fig:gal_hist}). In this test, we assume the GOODS-S field represents a blank field and an excess of galaxies relative to the GOODS-S in a magnitude bin would be an indication of contamination by cluster galaxies. The test found that our source catalogs are consistent with the number density in the GOODS-S catalog at bright magnitudes. For faint sources, our catalog has lower number density than the GOODS-S because of the difference in imaging depth. 3) We tested the impact that a redder source selection would have on the weak-lensing mass distribution and mass estimates. The low redshift of CIZAJ0107 means that many background galaxies are redder than the cluster red sequence. This is apparent in the color-magnitude diagrams of Figure \ref{fig:cmds}. A source catalog that contains only galaxies that are redder than the red sequence should be lower in contamination than our source catalog because it bypasses faint, blue cluster galaxies. We performed our weak-lensing analysis with a catalog of galaxies that were selected with the addition of a color constraint of $g-r > 1$ and F606W-F814W $ > 1$. We found that using this red galaxy catalog lowered the S/N of the peaks in the mass distribution and lowered the mass of each subcluster (within the statistical errors).}

Our source selection criteria provide 55.8 galaxies arcmin$^{-2}$ in the HST NE pointing, 51.4 galaxies arcmin$^{-2}$ in the HST SW pointing, and 9.8 galaxies arcmin$^{-2}$ in the HSC. These number densities are lower than we typically find in our weak-lensing analyses but are on par with the weak-lensing study of another Cluster in the Zone of Avoidance, CIZA J2242.8+5301 \citep{2014jee}. Primary causes of the low source number densities are the crowded star field and high extinction \citep[\mytilde1 magnitude in the R-band;][]{2011schlafly}. 

In addition to the selection of source galaxies, we selected candidate cluster member galaxies from the CMD. This was achieved by performing a linear fit to the spectroscopically confirmed cluster members. Subaru photometric candidates were then chosen to be galaxies within ($HSC$-$g-HSC$-$r2)\pm0.25$ of the linear fit and $16.5<HSC$-$r2<21$. For HST, the galaxies were selected within a color of (F606W-F814W$\pm0.1$) of a linear fit to the red sequence and $16.5<$F814W$<21$. To prevent stars from making their way into the catalog of cluster galaxies, we also constrained the FLUX\_RADIUS to be greater than 3.4 pixels (HSC) and 1.5 pixels (HST). Lastly, we removed any non-galaxy objects from the catalog by visual inspection of the color image. In total, we selected 212 photometric candidate cluster member galaxies in supplement to the \edit2{95} spectroscopic member galaxies.

\subsection{Source Redshift Determination} \label{sec:redshift_determination}
As shown by Equation \ref{eq:critical_density}, the weak-lensing signal depends on the distance to the background galaxies. Ideally, spectroscopic or photometric distance measurements would be available for each lensed galaxy. However, these measurements have not been obtained for the background galaxies in the CIZAJ0107 field. Instead, as past weak-lensing studies have done, a photometric redshift catalog can be used as a reference for the CIZAJ0107 field. For this study, we employ the GOODS-S photometric-redshift catalog \citep{2010dahlen} as a reference.

We constrained the GOODS-S catalog with the same photometric criteria as the source catalog (Section \ref{sec:source_selection}). This was done independently for the HST and HSC source catalogs because of the differing filter sets. \edit1{Figure \ref{fig:gal_hist} presents the number density of galaxies in magnitude bins for the GOODS-S catalog and our three source catalogs. When using the GOODS-S catalog as a reference, we weighted the GOODS-S catalog by the ratio of bins to make up for the difference in imaging depth.} The lensing efficiency of source galaxies is 
\begin{equation}
    \beta=\left<max(0,D_{ls}/D_s)\right>,
\end{equation}
where foreground galaxies ($D_{ls}/D_s<0$) are assigned zero. For the constrained reference catalogs, we find $\beta_{NE}=0.83$ and $\beta_{SW}=0.83$ for the NE and SW HST catalogs. For the HSC catalog, $\beta_{HSC}=0.79$. \edit2{Since foreground sources are assigned zero in the calculation of beta, the redshift that corresponds to beta is called an effective redshift ($z_{eff}$) rather than a mean redshift.} The effective redshifts for these $\beta$ values are $z_{eff,NE}=0.90$, $z_{eff,SW}=0.91$, and $z_{eff,HSC}=0.57$. Since we are representing the background galaxies with a single redshift, we must take the width of the distributions into account. The widths of each distribution are ${<}\beta_{NE}^2{>}=0.72$, ${<}\beta_{SW}^2{>}=0.72$, and ${<}\beta_{HSC}^2{>}=0.66$. These widths were applied to the reduced shear $g$ as follows \citep{1997seitz, 2000hoekstra}: 

\begin{equation}
    g' = \left[1 + \left(\frac{{<}\beta^2{>}}{{<}\beta{>}^2} - 1\right)\kappa\right]g,
\end{equation}
where $g'$ is the corrected reduced shear.

\begin{figure*}[!ht]
    \centering
    \includegraphics[width=\textwidth]{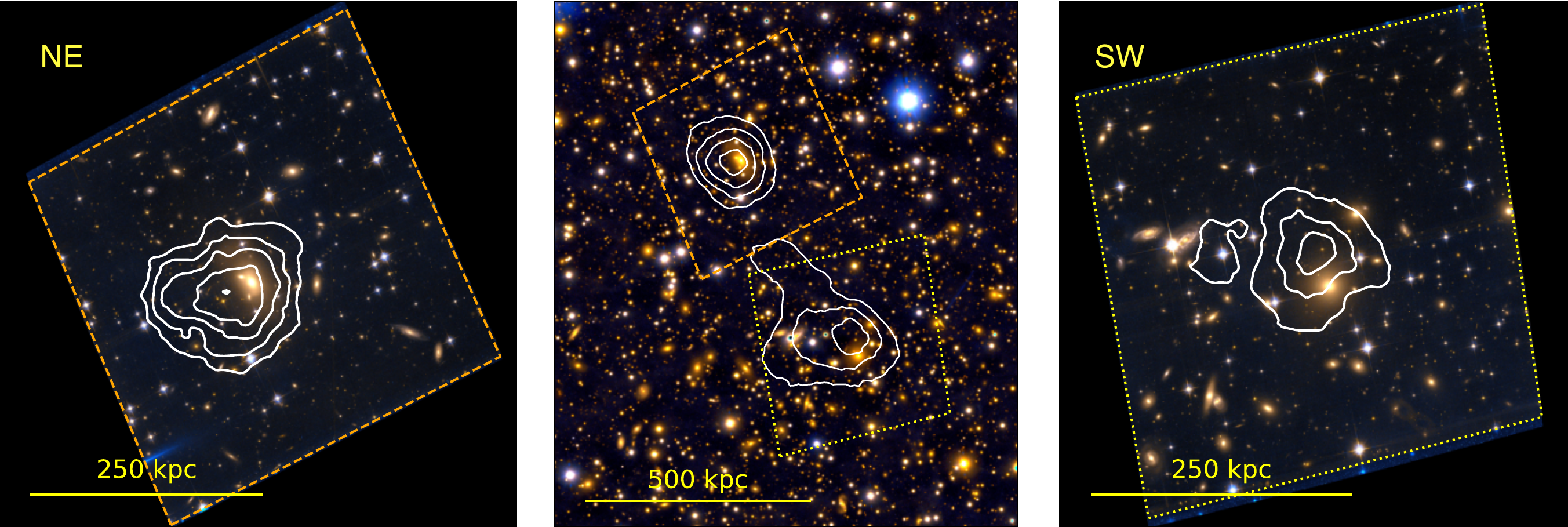}
    \caption{Mass distribution for CIZAJ0107. Contours display S/N of the lensing signal starting at $2\sigma$ and increasing in $1\sigma$ intervals. Left: NE HST color image with convergence contours. Middle: Subaru color image with convergence contours derived from a combined HST and Subaru shape catalog. Right: SW HST color image with convergence contours.}
    \label{fig:massmaps}
\end{figure*}

\section{Results}\label{sec:results}
\subsection{Mass Reconstruction}\label{sec:mass_reconstruction}
Weak lensing is an unparalleled tool for detecting the mass distribution of a galaxy cluster. The statistical nature of weak lensing requires an averaging of the source-galaxy shapes to detect the lensing signal. We utilize the FIATMAP code \citep{1997fischer} to produce weak-lensing mass maps. FIATMAP performs a real-space convolution to convert galaxy ellipticities to convergence.

HST-based mass maps of the NE and SW subclusters are shown in the left and right panels of Figure \ref{fig:massmaps}.  The convergence contours are chosen to start at a signal-to-noise (S/N) level of $2\sigma$ and increase at $1\sigma$ intervals. These S/N contours were determined by bootstrap resampling the source catalogs and creating 1000 realizations of the mass maps. The maximum S/N of the NE and SW mass distributions are $6.0\sigma$ and $4.6\sigma$, respectively. Neither weak-lensing peak is centered on its respective BCG with the NE peak offset $21^{+14}_{-7}$ kpc to the southeast of the BCG and the SW peak offset $27^{+19}_{-13}$ to the north of its BCG. These offsets are explored in Section \ref{sec:peaks} and agree with the respective BCG positions at the $2\sigma$ level. Qualitatively, the NE mass distribution (left panel) is round with no signs of substructure. The SW mass distribution (right panel) is round near the peak but, at lower significance, the contours stretch eastward towards two spectroscopically confirmed cluster spiral galaxies.  

The central panel exhibits the mass map derived from the combined HSC/HST source catalog. Again, the S/N contour levels begin at $2\sigma$ and increase in intervals of $1\sigma$. The peaks of the combined mass map are consistent with those of the individual HST mass maps. The NE mass distribution remains round and the west to east elongation of the SW weak-lensing signal persists but at a $3\sigma$ S/N level. There is also a tail, with a significance of $2\sigma$, that stretches along the axis connecting the subclusters. This tail passes over the X-ray emission peak (Figure \ref{fig:chandra}). The detection of this tail is intriguing but since the S/N of the detection is low, we will omit it from our subsequent analysis. 

\subsection{Mass Estimation}\label{sec:mass_estimation}
The mass of a galaxy cluster can be estimated through a number of methods. For weak lensing of a multi-cluster system, the proper method should simultaneously consider all significant subclusters. 

Our weak-lensing mass reconstruction shows that CIZAJ0107 is bimodal with two subclusters having weak-lensing signal above $3\sigma$. Taking the bimodality into consideration, we simultaneously fit two NFW halos \citep{1997navarro} to the projected mass distribution. This was accomplished by modeling the expected shear at each source galaxy location, following the NFW formalism presented in \cite{2000wright}, and minimizing the difference between the expected shear and the measured galaxy ellipticities (Section \ref{sec:psf_modeling}). To account for the degeneracy of mass and concentration, we used the mass-concentration relation of \cite{2019diemer}. The NFW halos were centered on their respective BCGs and source galaxies within 40 kpc of the BCGs were removed to minimize the bias caused be miscentering and to avoid regions where strong lensing could be present. To investigate the bias that is associated with the miscentering of the halo, we also performed a fit with the halos centered on their respective mass peaks. Centering on the mass peaks returned masses that are \mytilde3\% lower than centering on the BCGs. The maximum radius for fitting was chosen to be 20\arcmin\ (\mytilde2.3 Mpc) from the midpoint between the BCGs. This radius is \mytilde1.2 times larger than the expected $R_{200}$ of the cluster.  

\edit2{The background sources for CIZAJ0107 cover a wide distribution of redshifts because CIZAJ0107 is a low redshift cluster. Therefore, one must consider the variance that the large-scale structure introduces to the weak-lensing shear. We derived the convergence power spectrum for our background galaxy distribution using CAMB \citep{2011lewis}. We then calculate the expected covariance to the weak-lensing shear as a function of cluster-centric radius \citep[][]{2001hoekstra, 2003hoekstra, 2011umetsu}. The covariance of the LSS was added in quadrature to the statistical uncertainties and the mass of each subcluster was fit. The mass estimates are \edit2{$M_{200}=2.8^{+1.1}_{-1.1}\times10^{14}$ M$_\odot$} and \edit2{$M_{200}=3.1^{+1.2}_{-1.2}\times10^{14}$ M$_\odot$} for the NE and SW subclusters, respectively.} The mass estimates indicate a major merger of two approximately equal-mass clusters.

\section{Discussion}\label{sec:discussion}
\subsection{Mass of the System}
A primary input to modeling merging galaxy clusters is the mass of the subclusters and it is important to provide the most accurate mass estimates when initializing computer simulations. In Section \ref{sec:mass_estimation}, we estimated the mass of each subcluster. However, to compare with mass estimates from literature, we seek to understand the total mass of the cluster. One should not sum the subcluster masses, derived in Section \ref{sec:mass_estimation}, and compare them to the total mass estimates from literature because of the definition of $M_{200}$ and its dependence on $R_{200}$. 

To estimate the total weak-lensing mass of the system, we modeled two NFW halos on a uniform grid at a separation of 500 kpc (approximately the observed projected separation). From the barycenter of the system, we integrated radially until the average density within the sphere is 200 (500) times the critical density of the universe at $z=0.105$. We find the total weak-lensing mass of the system to be \edit2{$M_{200}=7.4^{+2.9}_{-2.9}\times10^{14} M_\odot$ ($M_{500}=5.3^{+2.1}_{-2.1}\times10^{14} M_\odot$)}.

%Our weak-lensing analysis found the SW subcluster mass to be larger than the NE subcluster mass. Qualitatively, this finding is in agreement with the X-ray features of the cluster. As is evident in the Bullet cluster, the mass of the bullet (western) subcluster is lower than the mass of the eastern cluster. CIZAJ0107 follows the same pattern. The southern X-ray emission is broad and round and is the location of the more massive subcluster. In contrast, the northern X-ray emission shows a triangular wake that trails towards the cluster center with the lower mass subcluster leading the wake feature.

CIZAJ0107 has few mass estimations performed. Planck estimations constrain the mass of the system to be $M_{500,SZ}=5.8\pm0.3\times10^{14} M_\odot$ \citep{2016planck}. Also based on the measurements from the ICM, \cite{2016randall} found the X-ray temperature to be $T=7.8\pm0.5$~keV, which gives $M_{500} = 7.8^{+0.8}_{-0.7}\times10^{14} M_\odot$ from the $M_{500}-T_X$ scaling relation \citep{2009vikhlinin}.  The total weak-lensing mass is consistent with the SZ mass and the X-ray mass.

\begin{figure}
    \centering
    \includegraphics[width=0.5\textwidth]{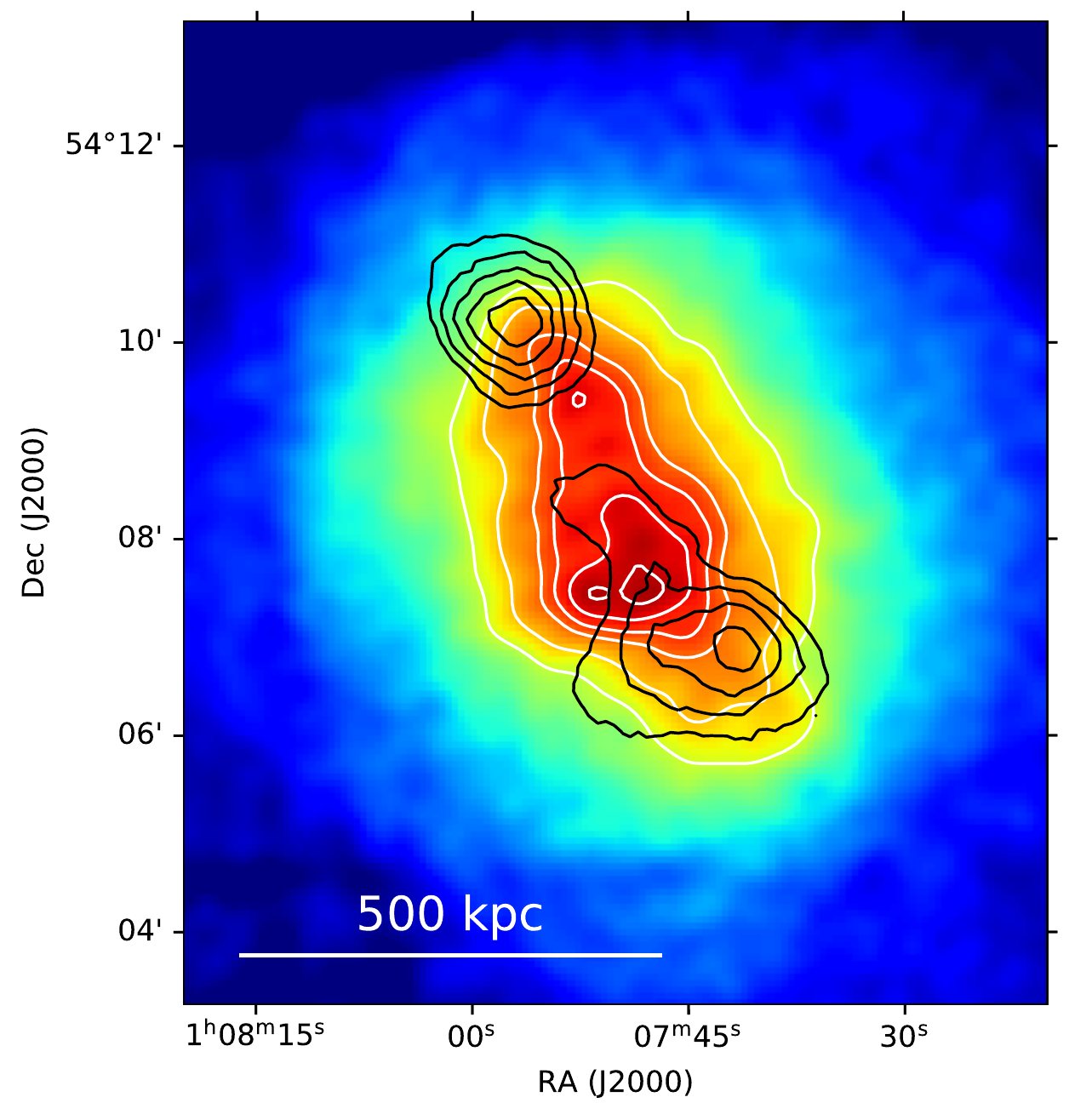}
    \caption{Weak lensing contours (black, from Figure \ref{fig:massmaps}) plotted over 160~ks, point-source-subtracted, exposure-corrected, 0.5-7~keV Chandra X-ray observation (Randall et al. in prep.). White contours highlight the X-ray emission. The Chandra image is smoothed with a $\sigma=8\arcsec$ circular Gaussian. The dissociative nature of CIZAJ0107 is clear with the weak-lensing peaks offset from the respective X-ray emission peaks.}
    \label{fig:chandra}
\end{figure}

\begin{figure}
    \centering
    \includegraphics[width=0.45\textwidth]{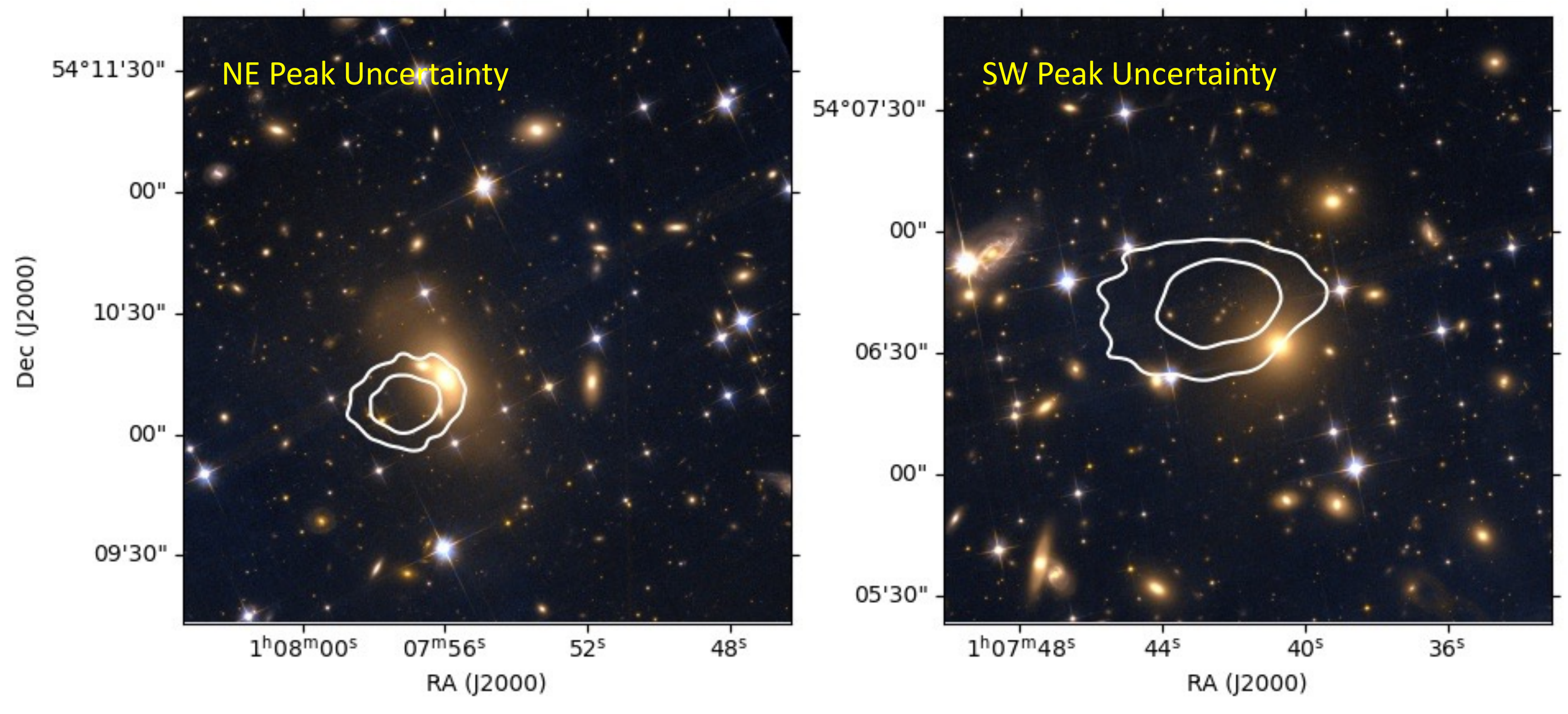}
    \caption{Uncertainty distribution of mass peak location derived from 1000 bootstrap samples of the source catalog. The inner contour contains $68\%$ and the outer contour contains $95\%$ of mass peaks. Left: NE subcluster with NE BCG in center. Right: SW subcluster with SW BCG in center.}
    \label{fig:centroids}
\end{figure}

\subsection{Mass Peak - BCG Offsets}\label{sec:peaks}
The large separation of the ICM X-ray brightness peaks from the weak-lensing signal (Figure \ref{fig:chandra}) indicates that CIZAJ0107 is a good candidate to probe a dark matter - BCG offset. Fortunately, the high density of source galaxies in the HST imaging permits tight constraints on the weak-lensing mass peaks.

Offsets of BCGs from mass peaks are common. In a weak-lensing analysis of galaxy clusters, \cite{2010oguri} found that the weak-lensing peak to BCG offset follows a Gaussian distribution with a width of $\sigma=90$ kpc. \cite{2012zitrin} put a tighter constraint on the offset of $\sigma=40$ kpc from strong lensing. The weak-lensing mass maps of CIZAJ0107 (Section \ref{sec:mass_reconstruction}) reveal offsets of the BCG from the mass peak of $21^{+14}_{-7}$ kpc and $27^{+19}_{-13}$ kpc in the NE and SW, respectively. These offsets are within the 1$\sigma$ ranges of \cite{2010oguri} and \cite{2012zitrin}.

To estimate the uncertainty on the weak-lensing mass peaks, we returned to the 1000 bootstrapped realizations of the mass maps and measured the weak-lensing peak location in each realization by elliptical Gaussian fitting. The centroid of the elliptical Gaussian fit for each realization was recorded and then the ensemble was passed to a Kernel Density Estimator to determine the $1\sigma$ and $2\sigma$ uncertainties of the peak location. Figure \ref{fig:centroids} displays the 1$\sigma$ and 2$\sigma$ uncertainty contours of the NE (left) and SW (right) subclusters. For both subclusters, the brightness peak of the BCG lies within the 2$\sigma$ range of the weak lensing mass peak.

\subsection{Merging Scenario}
With the wealth of multiwavelength observations, we can provide new insight into the past collision that CIZAJ0107 underwent. 

The weak-lensing analysis presented in this work highlights two prominent mass peaks that are found at opposite ends of the elongated X-ray distribution (Figure \ref{fig:chandra}). This weak-lensing signal traces the mass of the two subclusters that have recently collided. The SW subcluster is located closer to the SW (global) X-ray brightness peak than the NE subcluster is. The separation of the SW mass peak from the X-ray brightness peak is \mytilde160 kpc (1.35 arcmin). Furthermore, the SW mass peak is nestled between the X-ray brightness peak and the high temperature region that was highlighted in \cite{2016randall}. This configuration is consistent with the idea that the temperature boost is caused by a merger-induced shock that launched during the collision of the NE and SW subclusters. 

The NE mass peak is \mytilde160 kpc (1.35 arcmin) northeast of the NE X-ray peak and at the vertex of a roughly triangular-shaped X-ray emission. This appears to be ram-pressure stripping in action with a wake feature extending behind the deepest point of the gravitational potential. The alignment of the opening of the wake feature with the position of the SW subcluster and the symmetry of the wake feature may be an indication of a close to head-on collision. Also, the presence of the wake feature suggests that the merging system is likely viewed on its approach to first apocenter. We defer any further analysis of this feature to the upcoming X-ray analysis that includes new Chandra observations (Randall et al. in prep.).

In section \ref{sec:mmt}, we analyzed spectroscopic redshifts for the cluster galaxies and found that the BCGs have a \mytilde1200 km s$^{-1}$ difference in radial velocity. A two-component Gaussian fit returned two distributions with approximately the same radial velocity difference as the BCGs. The difference in radial velocity may indicate that our viewing angle of the merger is not perpendicular to the merger axis. However, it is difficult to say if the velocity difference of the BCGs is representative of the viewing angle of the past collision because the radial velocities at the closest approach (pericenter) can differ from the radial velocities when viewing the system near apocenter. It is likely that there is a line-of-sight component to the merger but perhaps not as significant as the BCG velocity difference is alluding to. \cite{2016randall} suggest that the lack of a well defined edge in the candidate SW radio relic could also be an indication of a line-of-sight component to the merger. 

The weak-lensing work presented here is the first result of new observations on CIZAJ0107. A new Chandra analysis is being performed with deeper observations. In addition, new VLA observations (Schwartzman et al. in preparation) and Low Frequency Array (LOFAR) observations will provide a better look into the radio structure of CIZAJ0107. In combination with our weak-lensing study, the multiwavelength data will provide improved constraining power on the merger scenario.

\begin{figure*}[!ht]
    \centering
    \includegraphics[width=\textwidth]{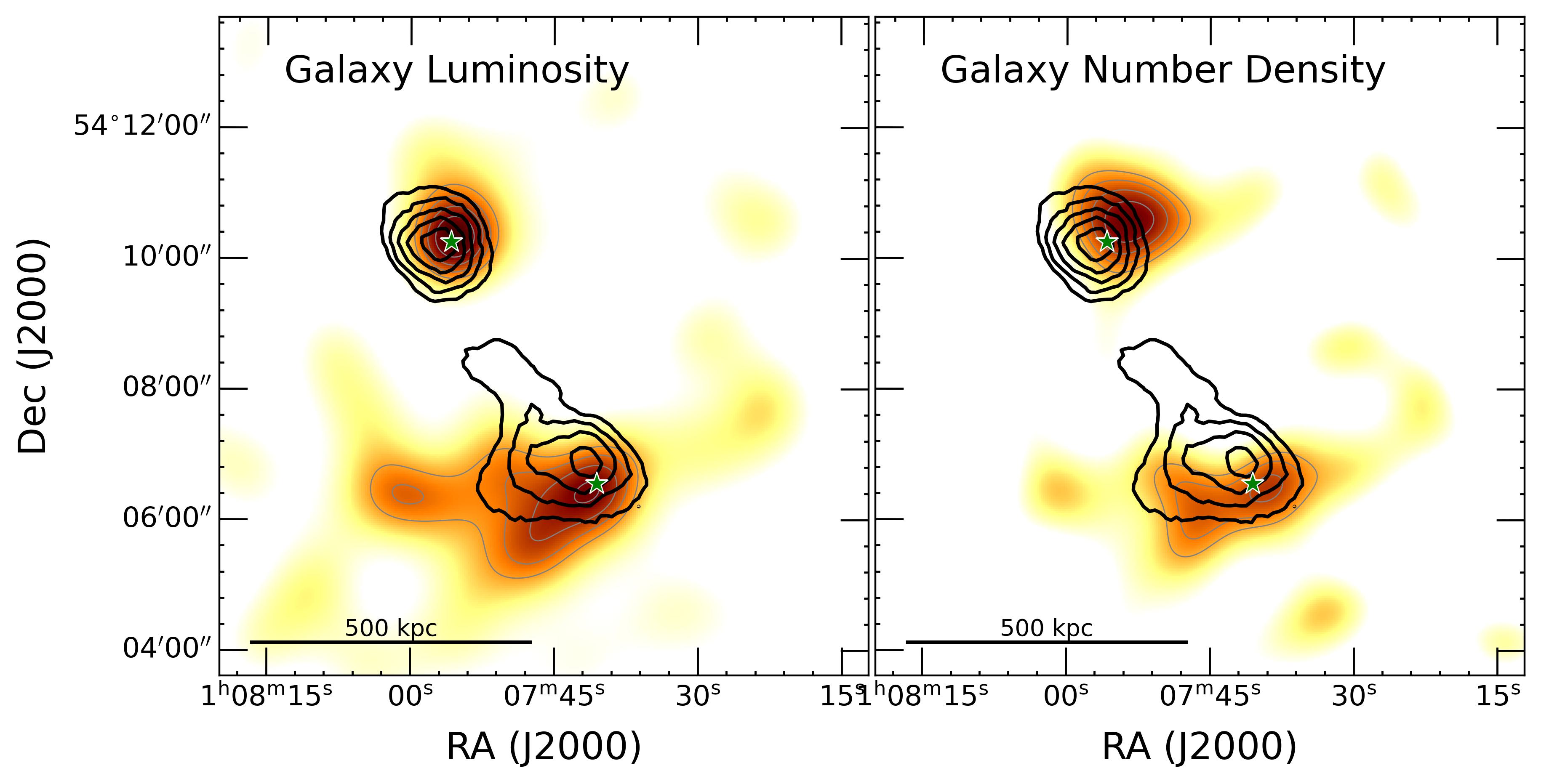}
    \caption{Weak-lensing mass contours (black; from Figure \ref{fig:massmaps}) over smoothed \textit{HSC-r2} galaxy luminosity (left) and smoothed galaxy number density (right). Grey contours highlight the colormaps. The galaxies used for plotting are a combination of spectroscopically confirmed and photometrically selected cluster members (Section \ref{sec:source_selection}). BCGs are marked with green stars. }
    \label{fig:galaxies}
\end{figure*}

\begin{figure}
    \centering
    \includegraphics[width=0.4\textwidth]{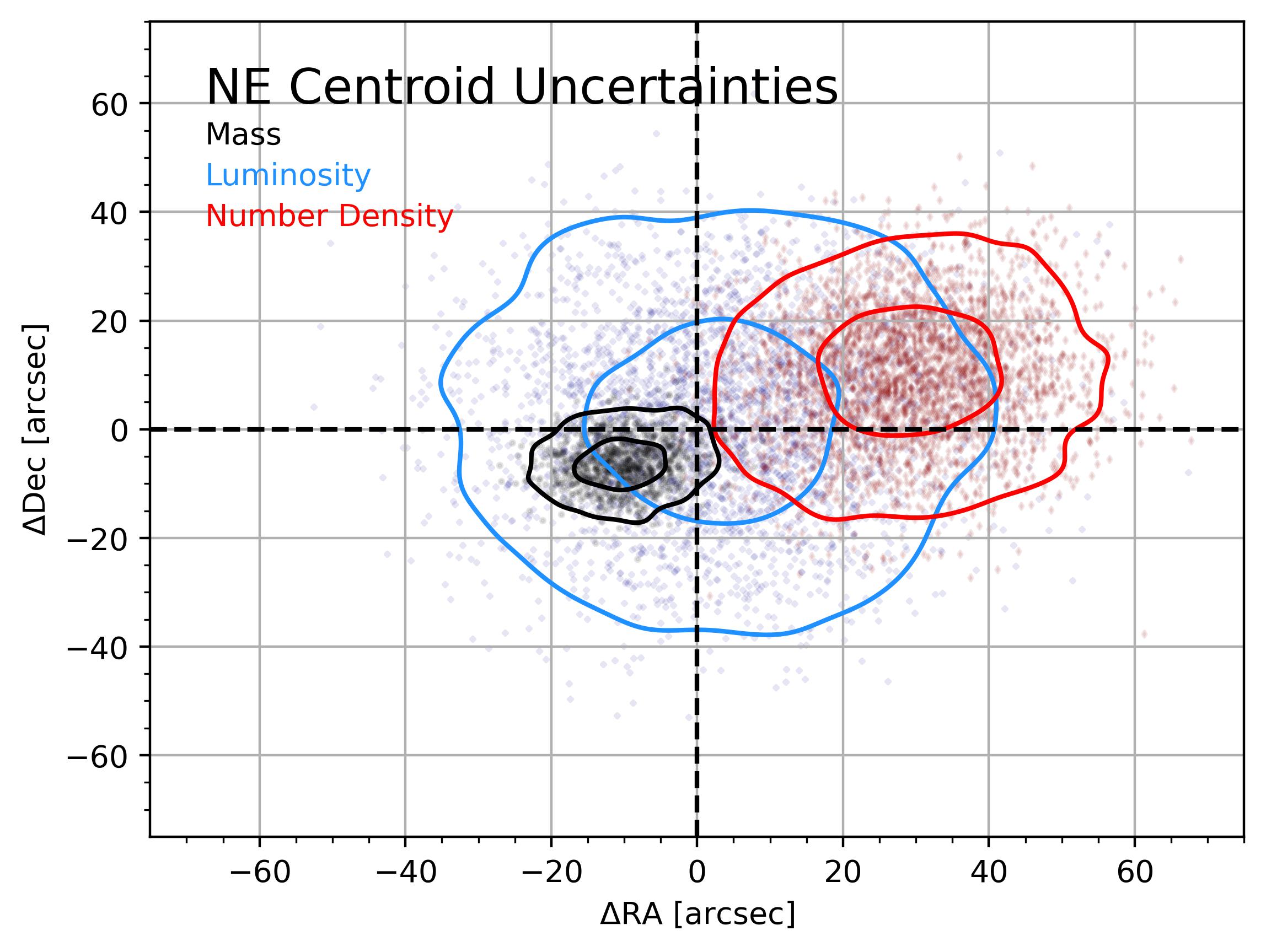}\\
    \includegraphics[width=0.4\textwidth]{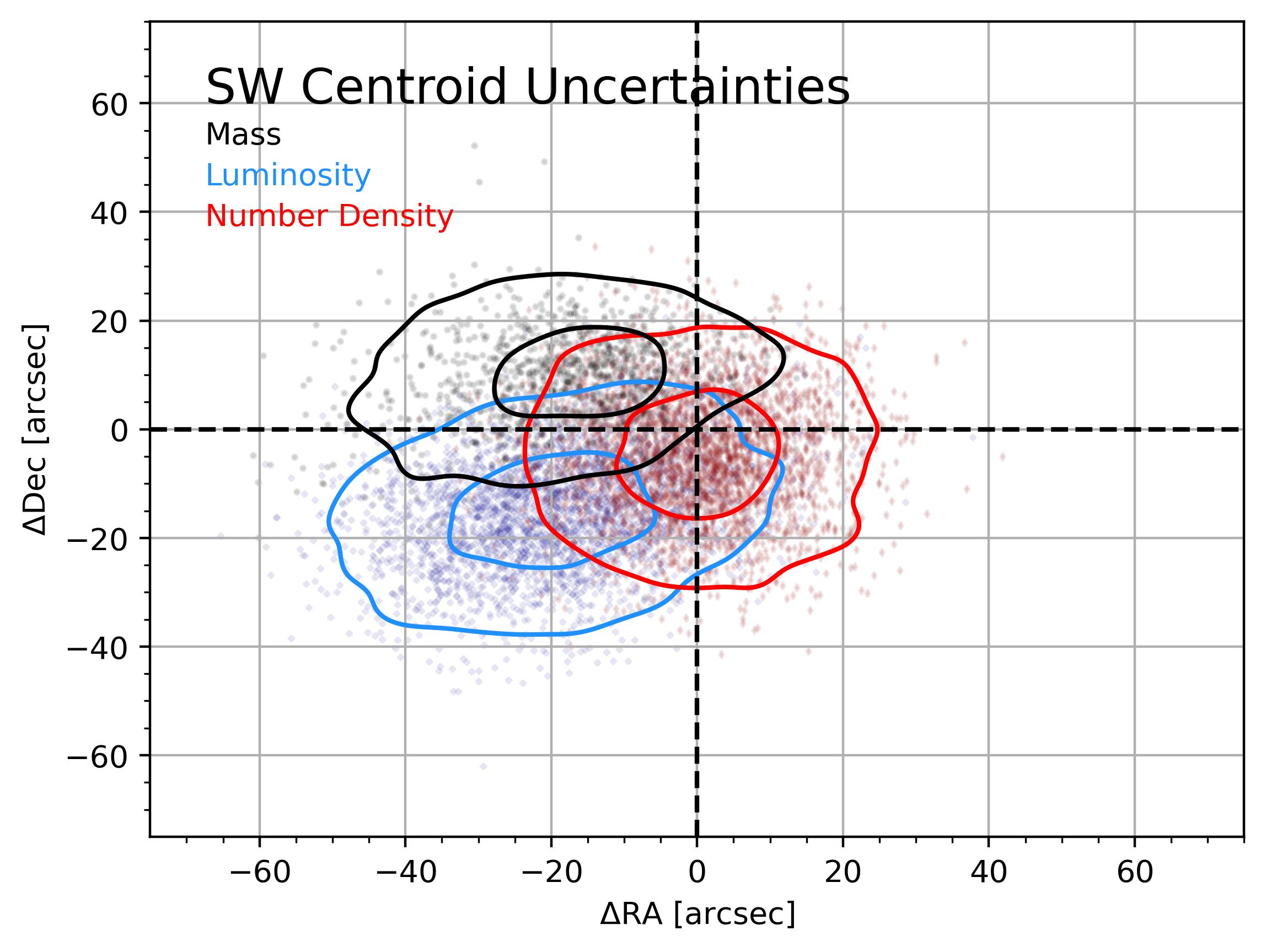}
    \caption{Centroid uncertainties for the NE (top) and SW (bottom) subclusters from bootstrap resampling. Contours represent the $1\sigma$ and $2\sigma$ uncertainties of the probability density function. The center (0,0) marks the position of the respective BCG. }
    \label{fig:peaks}
\end{figure}

\begin{figure}
    \centering
    \includegraphics[width=0.5\textwidth]{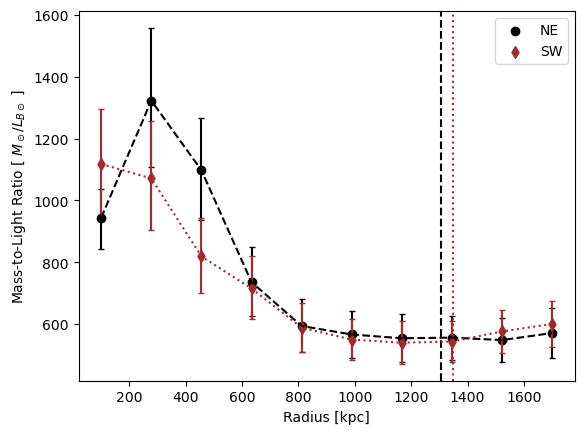}
    \caption{Cumulative mass-to-light ratio. Each profile is centered on its respective BCG. The vertical dashed (dotted) lines represent the virial radius of each subcluster from Section \ref{sec:mass_estimation}. }
    \label{fig:m2l}
\end{figure}

\subsection{Mass-galaxy Relation} \label{sec:mass_galaxy}
Separating the behavior of dark matter and galaxies is a key to detecting the self-interacting properties of dark matter. In \cite{2004markevitch}, three independent methods were used to put constraints on the cross-section of self-interacting dark matter (SIDM). Of these methods, the one with the best constraining power was a test of the mass-to-light ratio. In a cluster-cluster merger, the scattering of dark matter particles through self interaction could lead to a loss of dark matter for the halo   \citep{2004markevitch, 2008randall, 2017kim}. The reaction by the galaxies, caused by the gravitational force, follows later. Thus, observing the relation of galaxies and dark matter in merging clusters is a key diagnostic of SIDM. The nearly head-on, equal-mass, dissociative merger of CIZAJ0107 is an excellent target for investigating the properties of dark matter because the probability of self interactions depends on the local density of particles. Here, we compare the mass detected in weak lensing to the luminosity and number density of cluster galaxies.

Discerning the light emitted from all of the galaxies in a cluster is a difficult task. In Section \ref{sec:source_selection}, the spectroscopically confirmed cluster galaxies were combined with a photometrically selected sample of cluster galaxies to supplement the lacking spectroscopic coverage of CIZAJ0107. The left panel of Figure \ref{fig:galaxies} provides the weak-lensing mass map (black contours) plotted over the smoothed luminosity of cluster galaxies. In the north, the galaxy luminosity is compact with the majority of the light coming from the BCG and its companions. The weak-lensing signal in the north is also compact. On the other hand, the southern galaxy luminosity is dispersed over a large projected area with the luminosity peak stretching to the southeast of the BCG and the mass peak. The bright galaxies that stretch the luminosity peak to the southeast are visible in the right panel of Figure \ref{fig:massmaps}, with many galaxies of similar color located southeast of the BCG. One could interpret the elongated WL/starlight signal towards the southeast as a sign of a substructure but its existence is not separately detected in the X-ray emission. More HST observations may be required to discriminate a substructure in the weak-lensing map. The right panel of Figure \ref{fig:galaxies} shows the galaxy number density smoothed with the same Gaussian kernel as the luminosity. The number density of galaxies shows a slight shift from the luminosity peak in the north. In the south, the peak of the number density of galaxies is close to the BCG but has a distinct east-west elongation. It should be noted that the number density of galaxies is much more susceptible to incompleteness in the case of CIZAJ0107 because the crowded star field and extinction disproportionately affect the smaller and fainter cluster galaxies.

Figure \ref{fig:peaks} presents the $1\sigma$ and $2\sigma$ centroid uncertainty distributions for the luminosity of cluster galaxies (blue), number density of cluster galaxies (red), and weak-lensing signal (black). The luminosity and number density distributions were determined by calculating the first moment of 1000 bootstraps of the cluster galaxy catalog. The axes are defined relative to the centroid of the respective BCG. This figure quantifies what is qualitatively shown in Figure \ref{fig:galaxies}. In the NE, the galaxy luminosity (number density) centroid is in the $1\sigma$ ($2\sigma$) range of the mass peak. The BCG lies within the luminosity distribution and outside the number density distribution because this BCG does not have many nearby galaxies. In the SW, the galaxy luminosity (number density) centroid is in the $2\sigma$ ($1\sigma$) range of the mass peak. In this case, the BCG is not near the center of the luminosity distribution but instead within the number density distribution. As mentioned earlier, the bright galaxies to the southeast of the BCG have pulled the luminosity centroid towards them. Whereas, the number density centroid is dominated by the fainter companions to the BCG. The distance measurements from weak-lensing peak to respective galaxy centroids are presented in Table \ref{table:peak_offsets} with the $1\sigma$ errors from bootstrapping.

%\begin{table*}[]
%\caption {Peak Offsets from weak-lensing Peak} \label{table:peak_offsets}
%\def\arraystretch{1.}
%\begin{tabular}{ c  c  c c c }
%\hline
%\hline
%Peak & NE Physical & NE Angular & SW Physical & SW Angular           \\
% & [kpc] & [arcsec] & [kpc] & [arcsec]           \\
%\hline
%BCG                     & $18^{+15}_{-11}$  & $9^{+8}_{-6}$     & $28^{+25}_{-11}$    & $14^{+13}_{-6}$  \\
%Galaxy Luminosity       & $31^{+50}_{-31}$  & $16^{+26}_{-16}$  & $34^{+40}_{-34}$    & $18^{+21}_{-18}$ \\
%Galaxy Number Density   & $74^{+63}_{-19}$  & $38^{+33}_{-10}$  & $53^{+63}_{-27}$    & $27^{+23}_{-20}$ \\
%X-ray                   & 118           & 60        & 136          & 70 \\
%\hline

%\end{tabular}

%\end{table*}

\begin{table}[]
\caption {Peak Offsets from weak-lensing Peaks} \label{table:peak_offsets}
\begin{tabular}{ l  c  c  }
\hline
\hline
Peak & NE Subcluster  & SW Subcluster\\
 & [kpc] & [kpc]           \\
\hline
BCG                     & $21^{+14}_{-7}$       & $27^{+19}_{-13}$       \\
Luminosity       & $30^{+38}_{-30}$    & $40^{+46}_{-28}$     \\
Number Density   & $80^{+40}_{-33}$    & $41^{+33}_{-41}$     \\
%X-ray                   & 160           & 80        & 160          & 80 \\
\hline

\end{tabular}

\end{table}

\textbf{Mass-to-light:} In head-on collisions, a significant amount of dark matter may be scattered to large orbital radii, leading to a temporary drop of the $M/L$ ratio in the core and a boost in the outskirts \citep{2008randall, 2017kim}. Figure \ref{fig:m2l} shows the cumulative $M/L$ ratios for the NE and SW subclusters. The profiles were calculated by dividing the enclosed projected NFW mass by the enclosed cluster galaxy luminosity \citep[converted to B-band by synthetic photometry;][]{2005sirianni} within a given radius. The $M/L$ ratio of the northern subcluster initially increases with radius from the center because of the lack of galaxy light that surrounds the BCG (this is seen in Figure \ref{fig:galaxies}). This behavior has been observed in other clusters \citep{1997carlberg} and in cluster mergers such as Abell 115 \citep[e.g.][]{2019kim}. The profiles then decrease until $R_{200}$, where they flatten. At all radii, the $M/L$ ratios of the subclusters are statistically equal. The $M/L$ ratios within $R_{200}$ for the NE and SW subclusters are \edit2{$M/L=571^{+89}_{-91}\ M_\sun/L_{\sun,B}$} and \edit2{$M/L=564^{+87}_{-89}\ M_\sun/L_{\sun,B}$}, respectively. These $M/L$ ratios are in line with the expectation from other studies \citep{2004eke, 2015proctor}. Therefore, we see no substantial decrease in the M/L ratio that could arise for SIDM particles during a cluster merger. We also note that the statistically equal $M/L$ ratios are in agreement with the finding that CIZAJ0107 is a nearly equal-mass merger.  

%\cite{2017kim} simulated SIDM in head-on mergers of $10^{15}\ M_\sun$ subclusters. Their simulations show that DM scattering causes the $M/L$ ratio within 600 kpc of a subcluster to change by up to 20\% between first infall and second pericenter passage for the most extreme SIDM merger. It should be expected that this decrease in $M/L$ in the center leads to an increase in the $M/L$ in the outskirts. \cite{2017kim} predict that in the extreme case of $\sigma/m = 10 cm^2/g$ and a collision velocity \mytilde4300 km/s, about $50–60\%$ of the dark matter is expected to scatter out. This could lead to a drastic change in the $M/L$ ratio. Since the profiles in Figure \ref{fig:m2l} show decreasing $M/L$ ratio as a function of radius and are consistent between the two subclusters, we cannot make any conclusion for SIDM from the shapes of the profiles. Furthermore, the $M/L$ ratios within $R_{200}$ fit into the expectation from other studies \citep{2004eke, 2015proctor}.   

\section{Conclusions}
CIZAJ0107 belongs to a rare class of merging clusters, called dissociative mergers, that exhibit large offsets of their X-ray brightness peaks from their weak lensing and galaxy peaks. The special configuration of the mass components of dissociative mergers makes them important probes of dark matter and galaxy cluster formation theory. In this work, we combined the high-resolution imaging of the HST ACS with the wide field of the view of the Subaru HSC to thoroughly analyze the mass distribution of CIZAJ0107 with weak lensing, as a step towards better understanding dark matter.

Our weak-lensing analysis shows that the mass distribution of CIZAJ0107 is dominated by two subclusters at opposite ends of an elongated X-ray distribution. The weak-lensing signal of the NE subcluster is compact and round, as is its distribution of galaxies. In contrast, the weak-lensing signal of the SW subcluster is elongated in the west-east direction, which we interpret as a possible sign of substructure that is below the detection threshold of the available optical data. 

By simultaneously fitting the weak-lensing signal with two NFW halos, we found the mass of the NE subcluster to be \edit2{$M_{200}=2.8^{+1.1}_{-1.1}\times10^{14}$ M$_\odot$} and the mass of the SW subcluster to be \edit2{$M_{200}=3.1^{+1.2}_{-1.2}\times10^{14}$ M$_\odot$}. Based on the mass estimates for each halo, we derived the total mass of CIZAJ0107 to be \edit2{$M_{200}=7.4^{+2.9}_{-2.9}\times10^{14}$ M$_\odot$} and found it to be in the $1\sigma$ range of the SZ mass and the X-ray mass. The weak-lensing mass estimate of each subcluster will be integral in initializing future simulations of the CIZAJ0107 cluster merger.

We have taken the first steps to utilizing CIZAJ0107 for constraining the cross-section of SIDM. The locations of the mass peaks were compared to the locations of the X-ray peaks and the BCGs. The projected separation of the mass peaks from their closest X-ray brightness peaks were both measured to be \mytilde160 kpc. The offsets of the mass peak centroids from their respective BCGs are smaller ($21^{+14}_{-7}$ kpc in the NE and $27^{+19}_{-13}$ kpc in SW) than the X-ray offsets and found to be consistent at the $2\sigma$ level with their BCG brightness peak. 

As a test for the scattering of SIDM particles during the merger, we compared the $M/L$ ratios of each subcluster. Within $R_{200}$, the NE subcluster has \edit2{$M/L=571^{+89}_{-91}\ M_\sun/L_{\sun,B}$} and the SW has  \edit2{$M/L=564^{+87}_{-89}\ M_\sun/L_{\sun,B}$}. These $M/L$ ratio are found to be within the expectation for other galaxy clusters. We interpret the equality of the $M/L$ ratios as another indication of an equal-mass merger.

It is quite clear that CIZAJ0107 is an impressive example of a dissociative merger. Future radio, X-ray, and simulation results will further explore the merger details of CIZAJ0107.

\section{Acknowledgements}
We would like to thank the anonymous referee for comments that helped to improve the manuscript. We thank Tony Mroczkowski for useful discussions. This work is partly funded by NASA/Euclid grant 1484822. Partial support was also provided through an archival research grant from STScI under NASA contract NAS5-26555. Support for SWR was provided by the Chandra X-ray Center through NASA contract NAS8-03060, and the Smithsonian Institution. MJJ acknowledges support for the current research from the National Research Foundation of Korea under the program 2022R1A2C1003130. H.C. acknowledges support from the Brain Korea 21 FOUR Program. RJvW acknowledges support from the ERC Starting Grant ClusterWeb 804208. This work was supported by the K-GMT Science Program (PID: GEMINI-KR-2020B-011) of the Korea Astronomy and Space Science Institute (KASI). This research is based on observations made with the NASA/ESA Hubble Space Telescope obtained from the Space Telescope Science Institute, which is operated by the Association of Universities for Research in Astronomy, Inc., under NASA contract NAS 5–26555. These observations are associated with program 15610. Observations reported here were obtained at the MMT Observatory, a joint facility of the Smithsonian Institution and the University of Arizona. Based [in part] on data collected at the Subaru Telescope, via the time exchange program between Subaru and the international Gemini Observatory, a program of NSF’s NOIRLab. The Subaru Telescope is operated by the National Astronomical Observatory of Japan. We are honored and grateful for the opportunity of observing the Universe from Maunakea, which has the cultural, historical, and natural significance in Hawaii. Basic research in radio astronomy at the Naval Research Laboratory is supported by 6.1 base funding.

\software{Astrodrizzle \citep{2012gonzaga}, hscPipe \citep{2018bosch}, SCAMP \citep{bertinscamp}, SWarp \citep{bertinswarp}, SExtractor \citep{1996bertin}, scikit-learn \citep{scikit-learn}, NumPy \citep{numpy} }
\bibliography{refs}{}

\begin{thebibliography}{}
\expandafter\ifx\csname natexlab\endcsname\relax\def\natexlab#1{#1}\fi
\providecommand{\url}[1]{\href{#1}{#1}}
\providecommand{\dodoi}[1]{doi:~\href{http://doi.org/#1}{\nolinkurl{#1}}}
\providecommand{\doeprint}[1]{\href{http://ascl.net/#1}{\nolinkurl{http://ascl.net/#1}}}
\providecommand{\doarXiv}[1]{\href{https://arxiv.org/abs/#1}{\nolinkurl{https://arxiv.org/abs/#1}}}

\bibitem[{{Beers} {et~al.}(1990){Beers}, {Flynn}, \& {Gebhardt}}]{1990beers}
{Beers}, T.~C., {Flynn}, K., \& {Gebhardt}, K. 1990, \aj, 100, 32,
  \dodoi{10.1086/115487}

\bibitem[{{Bertin}(2006)}]{bertinscamp}
{Bertin}, E. 2006, in Astronomical Society of the Pacific Conference Series,
  Vol. 351, Astronomical Data Analysis Software and Systems XV, ed.
  C.~{Gabriel}, C.~{Arviset}, D.~{Ponz}, \& S.~{Enrique}, 112

\bibitem[{{Bertin} \& {Arnouts}(1996)}]{1996bertin}
{Bertin}, E., \& {Arnouts}, S. 1996, \aaps, 117, 393,
  \dodoi{10.1051/aas:1996164}

\bibitem[{{Bertin} {et~al.}(2002){Bertin}, {Mellier}, {Radovich}, {Missonnier},
  {Didelon}, \& {Morin}}]{bertinswarp}
{Bertin}, E., {Mellier}, Y., {Radovich}, M., {et~al.} 2002, in Astronomical
  Society of the Pacific Conference Series, Vol. 281, Astronomical Data
  Analysis Software and Systems XI, ed. D.~A. {Bohlender}, D.~{Durand}, \&
  T.~H. {Handley}, 228

\bibitem[{{Bosch} {et~al.}(2018){Bosch}, {Armstrong}, {Bickerton}, {Furusawa},
  {Ikeda}, {Koike}, {Lupton}, {Mineo}, {Price}, {Takata}, {Tanaka}, {Yasuda},
  {AlSayyad}, {Becker}, {Coulton}, {Coupon}, {Garmilla}, {Huang}, {Krughoff},
  {Lang}, {Leauthaud}, {Lim}, {Lust}, {MacArthur}, {Mandelbaum}, {Miyatake},
  {Miyazaki}, {Murata}, {More}, {Okura}, {Owen}, {Swinbank}, {Strauss},
  {Yamada}, \& {Yamanoi}}]{2018bosch}
{Bosch}, J., {Armstrong}, R., {Bickerton}, S., {et~al.} 2018, \pasj, 70, S5,
  \dodoi{10.1093/pasj/psx080}

\bibitem[{{Brada{\v{c}}} {et~al.}(2008){Brada{\v{c}}}, {Allen}, {Treu},
  {Ebeling}, {Massey}, {Morris}, {von der Linden}, \& {Applegate}}]{2008bradac}
{Brada{\v{c}}}, M., {Allen}, S.~W., {Treu}, T., {et~al.} 2008, \apj, 687, 959,
  \dodoi{10.1086/591246}

\bibitem[{{Burkert}(2000)}]{2000burkert}
{Burkert}, A. 2000, \apjl, 534, L143, \dodoi{10.1086/312674}

\bibitem[{{Carlberg} {et~al.}(1997){Carlberg}, {Yee}, \&
  {Ellingson}}]{1997carlberg}
{Carlberg}, R.~G., {Yee}, H.~K.~C., \& {Ellingson}, E. 1997, \apj, 478, 462,
  \dodoi{10.1086/303805}

\bibitem[{{Cho} {et~al.}(2022){Cho}, {James Jee}, {Smith}, {Finner}, \&
  {Lee}}]{2022Cho}
{Cho}, H., {James Jee}, M., {Smith}, R., {Finner}, K., \& {Lee}, W. 2022, \apj,
  925, 68, \dodoi{10.3847/1538-4357/ac36c8}

\bibitem[{{Clowe} {et~al.}(2006){Clowe}, {Brada{\v{c}}}, {Gonzalez},
  {Markevitch}, {Randall}, {Jones}, \& {Zaritsky}}]{2006clowe}
{Clowe}, D., {Brada{\v{c}}}, M., {Gonzalez}, A.~H., {et~al.} 2006, \apjl, 648,
  L109, \dodoi{10.1086/508162}

\bibitem[{Dahlen {et~al.}(2010)Dahlen, Mobasher, Dickinson, Ferguson,
  Giavalisco, Grogin, Guo, Koekemoer, Lee, Lee, Nonino, Riess, \&
  Salimbeni}]{2010dahlen}
Dahlen, T., Mobasher, B., Dickinson, M., {et~al.} 2010, The Astrophysical
  Journal, 724, 425.
\newblock \url{http://stacks.iop.org/0004-637X/724/i=1/a=425}

\bibitem[{{Dawson} {et~al.}(2012){Dawson}, {Wittman}, {Jee}, {Gee}, {Hughes},
  {Tyson}, {Schmidt}, {Thorman}, {Brada{\v{c}}}, {Miyazaki}, {Lemaux},
  {Utsumi}, \& {Margoniner}}]{2012dawson}
{Dawson}, W.~A., {Wittman}, D., {Jee}, M.~J., {et~al.} 2012, \apjl, 747, L42,
  \dodoi{10.1088/2041-8205/747/2/L42}

\bibitem[{{Diemer} \& {Joyce}(2019)}]{2019diemer}
{Diemer}, B., \& {Joyce}, M. 2019, \apj, 871, 168,
  \dodoi{10.3847/1538-4357/aafad6}

\bibitem[{{Ebeling} {et~al.}(2002){Ebeling}, {Mullis}, \&
  {Tully}}]{2002ebeling}
{Ebeling}, H., {Mullis}, C.~R., \& {Tully}, R.~B. 2002, \apj, 580, 774,
  \dodoi{10.1086/343790}

\bibitem[{{Eke} {et~al.}(2004){Eke}, {Frenk}, {Baugh}, {Cole}, {Norberg},
  {Peacock}, {Baldry}, {Bland-Hawthorn}, {Bridges}, {Cannon}, {Colless},
  {Collins}, {Couch}, {Dalton}, {de Propris}, {Driver}, {Efstathiou}, {Ellis},
  {Glazebrook}, {Jackson}, {Lahav}, {Lewis}, {Lumsden}, {Maddox}, {Madgwick},
  {Peterson}, {Sutherland}, \& {Taylor}}]{2004eke}
{Eke}, V.~R., {Frenk}, C.~S., {Baugh}, C.~M., {et~al.} 2004, \mnras, 355, 769,
  \dodoi{10.1111/j.1365-2966.2004.08354.x}

\bibitem[{{Finner} {et~al.}(2020){Finner}, {James Jee}, {Webb}, {Wilson},
  {Perlmutter}, {Muzzin}, \& {Hlavacek-Larrondo}}]{2020finner}
{Finner}, K., {James Jee}, M., {Webb}, T., {et~al.} 2020, \apj, 893, 10,
  \dodoi{10.3847/1538-4357/ab7bdb}

\bibitem[{{Finner} {et~al.}(2017){Finner}, {Jee}, {Golovich}, {Wittman},
  {Dawson}, {Gruen}, {Koekemoer}, {Lemaux}, \& {Seitz}}]{2017finner}
{Finner}, K., {Jee}, M.~J., {Golovich}, N., {et~al.} 2017, \apj, 851, 46,
  \dodoi{10.3847/1538-4357/aa998c}

\bibitem[{{Finner} {et~al.}(2021){Finner}, {HyeongHan}, {Jee}, {Wittman},
  {Forman}, {van Weeren}, {Golovich}, {Dawson}, {Jones}, {de Gasperin}, \&
  {Jones}}]{2021finner}
{Finner}, K., {HyeongHan}, K., {Jee}, M.~J., {et~al.} 2021, \apj, 918, 72,
  \dodoi{10.3847/1538-4357/ac0d00}

\bibitem[{{Fischer} \& {Tyson}(1997)}]{1997fischer}
{Fischer}, P., \& {Tyson}, J.~A. 1997, \aj, 114, 14, \dodoi{10.1086/118447}

\bibitem[{{Flewelling} {et~al.}(2020){Flewelling}, {Magnier}, {Chambers},
  {Heasley}, {Holmberg}, {Huber}, {Sweeney}, {Waters}, {Calamida}, {Casertano},
  {Chen}, {Farrow}, {Hasinger}, {Henderson}, {Long}, {Metcalfe}, {Narayan},
  {Nieto-Santisteban}, {Norberg}, {Rest}, {Saglia}, {Szalay}, {Thakar},
  {Tonry}, {Valenti}, {Werner}, {White}, {Denneau}, {Draper}, {Hodapp},
  {Jedicke}, {Kaiser}, {Kudritzki}, {Price}, {Wainscoat}, {Chastel}, {McLean},
  {Postman}, \& {Shiao}}]{2020panstarrs}
{Flewelling}, H.~A., {Magnier}, E.~A., {Chambers}, K.~C., {et~al.} 2020, \apjs,
  251, 7, \dodoi{10.3847/1538-4365/abb82d}

\bibitem[{{Golovich} {et~al.}(2019){Golovich}, {Dawson}, {Wittman}, {van
  Weeren}, {Andrade-Santos}, {Jee}, {Benson}, {de Gasperin}, {Venturi},
  {Bonafede}, {Sobral}, {Ogrean}, {Lemaux}, {Brada{\v{c}}}, {Br{\"u}ggen}, \&
  {Peter}}]{2019bgolovich}
{Golovich}, N., {Dawson}, W.~A., {Wittman}, D.~M., {et~al.} 2019, \apj, 882,
  69, \dodoi{10.3847/1538-4357/ab2f90}

\bibitem[{{Gonzaga} {et~al.}(2012){Gonzaga}, {Hack}, {Fruchter}, \&
  {Mack}}]{2012gonzaga}
{Gonzaga}, S., {Hack}, W., {Fruchter}, A., \& {Mack}, J. 2012, {The DrizzlePac
  Handbook}

\bibitem[{Harris {et~al.}(2020)Harris, Millman, van~der Walt, Gommers,
  Virtanen, Cournapeau, Wieser, Taylor, Berg, Smith, Kern, Picus, Hoyer, van
  Kerkwijk, Brett, Haldane, del R{\'{i}}o, Wiebe, Peterson,
  G{\'{e}}rard-Marchant, Sheppard, Reddy, Weckesser, Abbasi, Gohlke, \&
  Oliphant}]{numpy}
Harris, C.~R., Millman, K.~J., van~der Walt, S.~J., {et~al.} 2020, Nature, 585,
  357, \dodoi{10.1038/s41586-020-2649-2}

\bibitem[{{Hoekstra}(2001)}]{2001hoekstra}
{Hoekstra}, H. 2001, \aap, 370, 743, \dodoi{10.1051/0004-6361:20010293}

\bibitem[{{Hoekstra}(2003)}]{2003hoekstra}
---. 2003, \mnras, 339, 1155, \dodoi{10.1046/j.1365-8711.2003.06264.x}

\bibitem[{{Hoekstra} {et~al.}(2000){Hoekstra}, {Franx}, \&
  {Kuijken}}]{2000hoekstra}
{Hoekstra}, H., {Franx}, M., \& {Kuijken}, K. 2000, \apj, 532, 88,
  \dodoi{10.1086/308556}

\bibitem[{{HyeongHan} {et~al.}(2020){HyeongHan}, {Jee}, {Rudnick}, {Parkinson},
  {Finner}, {Yoon}, {Lee}, {Brunetti}, {Br{\"u}ggen}, {Collier}, {Hopkins},
  {Micha{\l}owski}, {Norris}, \& {Riseley}}]{2020hyeonghan}
{HyeongHan}, K., {Jee}, M.~J., {Rudnick}, L., {et~al.} 2020, \apj, 900, 127,
  \dodoi{10.3847/1538-4357/aba742}

\bibitem[{{Jee} {et~al.}(2007){Jee}, {Blakeslee}, {Sirianni}, {Martel},
  {White}, \& {Ford}}]{2007jee}
{Jee}, M.~J., {Blakeslee}, J.~P., {Sirianni}, M., {et~al.} 2007, \pasp, 119,
  1403, \dodoi{10.1086/524849}

\bibitem[{{Jee} {et~al.}(2016){Jee}, {Dawson}, {Stroe}, {Wittman}, {van
  Weeren}, {Br{\"u}ggen}, {Brada{\v{c}}}, \& {R{\"o}ttgering}}]{2016jee}
{Jee}, M.~J., {Dawson}, W.~A., {Stroe}, A., {et~al.} 2016, \apj, 817, 179,
  \dodoi{10.3847/0004-637X/817/2/179}

\bibitem[{{Jee} {et~al.}(2014){Jee}, {Hughes}, {Menanteau}, {Sif{\'o}n},
  {Mandelbaum}, {Barrientos}, {Infante}, \& {Ng}}]{2014jee}
{Jee}, M.~J., {Hughes}, J.~P., {Menanteau}, F., {et~al.} 2014, \apj, 785, 20,
  \dodoi{10.1088/0004-637X/785/1/20}

\bibitem[{{Jee} \& {Tyson}(2011)}]{2011jeeLSST}
{Jee}, M.~J., \& {Tyson}, J.~A. 2011, \pasp, 123, 596, \dodoi{10.1086/660137}

\bibitem[{{Jee} {et~al.}(2013){Jee}, {Tyson}, {Schneider}, {Wittman},
  {Schmidt}, \& {Hilbert}}]{2013jee}
{Jee}, M.~J., {Tyson}, J.~A., {Schneider}, M.~D., {et~al.} 2013, \apj, 765, 74,
  \dodoi{10.1088/0004-637X/765/1/74}

\bibitem[{{Jee} {et~al.}(2011){Jee}, {Dawson}, {Hoekstra}, {Perlmutter},
  {Rosati}, {Brodwin}, {Suzuki}, {Koester}, {Postman}, {Lubin}, {Meyers},
  {Stanford}, {Barbary}, {Barrientos}, {Eisenhardt}, {Ford}, {Gilbank},
  {Gladders}, {Gonzalez}, {Harris}, {Huang}, {Lidman}, {Rykoff}, {Rubin}, \&
  {Spadafora}}]{2011jee}
{Jee}, M.~J., {Dawson}, K.~S., {Hoekstra}, H., {et~al.} 2011, \apj, 737, 59,
  \dodoi{10.1088/0004-637X/737/2/59}

\bibitem[{{Jee} {et~al.}(2015){Jee}, {Stroe}, {Dawson}, {Wittman}, {Hoekstra},
  {Br{\"u}ggen}, {R{\"o}ttgering}, {Sobral}, \& {van Weeren}}]{2015jee}
{Jee}, M.~J., {Stroe}, A., {Dawson}, W., {et~al.} 2015, \apj, 802, 46,
  \dodoi{10.1088/0004-637X/802/1/46}

\bibitem[{{Kim} {et~al.}(2019){Kim}, {Jee}, {Finner}, {Golovich}, {Wittman},
  {van Weeren}, \& {Dawson}}]{2019kim}
{Kim}, M., {Jee}, M.~J., {Finner}, K., {et~al.} 2019, \apj, 874, 143,
  \dodoi{10.3847/1538-4357/ab0d7c}

\bibitem[{{Kim} {et~al.}(2017){Kim}, {Peter}, \& {Wittman}}]{2017kim}
{Kim}, S.~Y., {Peter}, A. H.~G., \& {Wittman}, D. 2017, \mnras, 469, 1414,
  \dodoi{10.1093/mnras/stx896}

\bibitem[{{Kurtz} \& {Mink}(1998)}]{1998kurtz}
{Kurtz}, M.~J., \& {Mink}, D.~J. 1998, \pasp, 110, 934, \dodoi{10.1086/316207}

\bibitem[{{Lewis} \& {Challinor}(2011)}]{2011lewis}
{Lewis}, A., \& {Challinor}, A. 2011, {CAMB: Code for Anisotropies in the
  Microwave Background}, Astrophysics Source Code Library, record
  ascl:1102.026.
\newblock \doeprint{1102.026}

\bibitem[{{Markevitch} {et~al.}(2004){Markevitch}, {Gonzalez}, {Clowe},
  {Vikhlinin}, {Forman}, {Jones}, {Murray}, \& {Tucker}}]{2004markevitch}
{Markevitch}, M., {Gonzalez}, A.~H., {Clowe}, D., {et~al.} 2004, \apj, 606,
  819, \dodoi{10.1086/383178}

\bibitem[{{Markwardt}(2009)}]{2009markwardt}
{Markwardt}, C.~B. 2009, Astronomical Society of the Pacific Conference Series,
  Vol. 411, {Non-linear Least-squares Fitting in IDL with MPFIT}, ed. D.~A.
  {Bohlender}, D.~{Durand}, \& P.~{Dowler}, 251

\bibitem[{{Navarro} {et~al.}(1997){Navarro}, {Frenk}, \& {White}}]{1997navarro}
{Navarro}, J.~F., {Frenk}, C.~S., \& {White}, S. D.~M. 1997, \apj, 490, 493,
  \dodoi{10.1086/304888}

\bibitem[{{Oguri} {et~al.}(2010){Oguri}, {Takada}, {Okabe}, \&
  {Smith}}]{2010oguri}
{Oguri}, M., {Takada}, M., {Okabe}, N., \& {Smith}, G.~P. 2010, \mnras, 405,
  2215, \dodoi{10.1111/j.1365-2966.2010.16622.x}

\bibitem[{{Okabe} {et~al.}(2010){Okabe}, {Takada}, {Umetsu}, {Futamase}, \&
  {Smith}}]{2010okabe}
{Okabe}, N., {Takada}, M., {Umetsu}, K., {Futamase}, T., \& {Smith}, G.~P.
  2010, \pasj, 62, 811, \dodoi{10.1093/pasj/62.3.811}

\bibitem[{Pedregosa {et~al.}(2011)Pedregosa, Varoquaux, Gramfort, Michel,
  Thirion, Grisel, Blondel, Prettenhofer, Weiss, Dubourg, Vanderplas, Passos,
  Cournapeau, Brucher, Perrot, \& Duchesnay}]{scikit-learn}
Pedregosa, F., Varoquaux, G., Gramfort, A., {et~al.} 2011, Journal of Machine
  Learning Research, 12, 2825

\bibitem[{{Planck Collaboration} {et~al.}(2014){Planck Collaboration}, {Ade},
  {Aghanim}, {Armitage-Caplan}, {Arnaud}, {Ashdown}, {Atrio-Barand ela},
  {Aumont}, {Aussel}, {Baccigalupi}, {Band ay}, {Barreiro}, {Barrena},
  {Bartelmann}, {Bartlett}, {Battaner}, {Benabed}, {Beno{\^\i}t},
  {Benoit-L{\'e}vy}, {Bernard}, {Bersanelli}, {Bielewicz}, {Bikmaev}, {Bobin},
  {Bock}, {B{\"o}hringer}, {Bonaldi}, {Bond}, {Borrill}, {Bouchet}, {Bridges},
  {Bucher}, {Burenin}, {Burigana}, {Butler}, {Cardoso}, {Carvalho}, {Catalano},
  {Challinor}, {Chamballu}, {Chary}, {Chen}, {Chiang}, {Chiang}, {Chon},
  {Christensen}, {Churazov}, {Church}, {Clements}, {Colombi}, {Colombo},
  {Comis}, {Couchot}, {Coulais}, {Crill}, {Curto}, {Cuttaia}, {Da Silva},
  {Dahle}, {Danese}, {Davies}, {Davis}, {de Bernardis}, {de Rosa}, {de Zotti},
  {Delabrouille}, {Delouis}, {D{\'e}mocl{\`e}s}, {D{\'e}sert}, {Dickinson},
  {Diego}, {Dolag}, {Dole}, {Donzelli}, {Dor{\'e}}, {Douspis}, {Dupac},
  {Efstathiou}, {Eisenhardt}, {En{\ss}lin}, {Eriksen}, {Feroz}, {Finelli},
  {Flores-Cacho}, {Forni}, {Frailis}, {Franceschi}, {Fromenteau}, {Galeotta},
  {Ganga}, {G{\'e}nova-Santos}, {Giard}, {Giardino}, {Gilfanov},
  {Giraud-H{\'e}raud}, {Gonz{\'a}lez-Nuevo}, {G{\'o}rski}, {Grainge},
  {Gratton}, {Gregorio}, {Groeneboom}, {Gruppuso}, {Hansen}, {Hanson},
  {Harrison}, {Hempel}, {Henrot-Versill{\'e}}, {Hern{\'a}ndez-Monteagudo},
  {Herranz}, {Hildebrandt}, {Hivon}, {Hobson}, {Holmes}, {Hornstrup}, {Hovest},
  {Huffenberger}, {Hurier}, {Hurley-Walker}, {Jaffe}, {Jaffe}, {Jones},
  {Juvela}, {Keih{\"a}nen}, {Keskitalo}, {Khamitov}, {Kisner}, {Kneissl},
  {Knoche}, {Knox}, {Kunz}, {Kurki-Suonio}, {Lagache}, {L{\"a}hteenm{\"a}ki},
  {Lamarre}, {Lasenby}, {Laureijs}, {Lawrence}, {Leahy}, {Leonardi},
  {Le{\'o}n-Tavares}, {Lesgourgues}, {Li}, {Liddle}, {Liguori}, {Lilje},
  {Linden-V{\o}rnle}, {L{\'o}pez-Caniego}, {Lubin}, {Mac{\'\i}as-P{\'e}rez},
  {MacTavish}, {Maffei}, {Maino}, {Mandolesi}, {Maris}, {Marshall}, {Martin},
  {Mart{\'\i}nez-Gonz{\'a}lez}, {Masi}, {Massardi}, {Matarrese}, {Matthai},
  {Mazzotta}, {Mei}, {Meinhold}, {Melchiorri}, {Melin}, {Mendes}, {Mennella},
  {Migliaccio}, {Mikkelsen}, {Mitra}, {Miville-Desch{\^e}nes}, {Moneti},
  {Montier}, {Morgante}, {Mortlock}, {Munshi}, {Murphy}, {Naselsky}, {Nati},
  {Natoli}, {Nesvadba}, {Netterfield}, {N{\o}rgaard-Nielsen}, {Noviello},
  {Novikov}, {Novikov}, {O'Dwyer}, {Olamaie}, {Osborne}, {Oxborrow}, {Paci},
  {Pagano}, {Pajot}, {Paoletti}, {Pasian}, {Patanchon}, {Pearson}, {Perdereau},
  {Perotto}, {Perrott}, {Perrotta}, {Piacentini}, {Piat}, {Pierpaoli},
  {Pietrobon}, {Plaszczynski}, {Pointecouteau}, {Polenta}, {Ponthieu}, {Popa},
  {Poutanen}, {Pratt}, {Pr{\'e}zeau}, {Prunet}, {Puget}, {Rachen}, {Reach},
  {Rebolo}, {Reinecke}, {Remazeilles}, {Renault}, {Ricciardi}, {Riller},
  {Ristorcelli}, {Rocha}, {Rosset}, {Roudier}, {Rowan-Robinson},
  {Rubi{\~n}o-Mart{\'\i}n}, {Rumsey}, {Rusholme}, {Sandri}, {Santos},
  {Saunders}, {Savini}, {Schammel}, {Scott}, {Seiffert}, {Shellard},
  {Shimwell}, {Spencer}, {Stanford}, {Starck}, {Stolyarov}, {Stompor},
  {Sudiwala}, {Sunyaev}, {Sureau}, {Sutton}, {Suur-Uski}, {Sygnet}, {Tauber},
  {Tavagnacco}, {Terenzi}, {Toffolatti}, {Tomasi}, {Tristram}, {Tucci},
  {Tuovinen}, {T{\"u}rler}, {Umana}, {Valenziano}, {Valiviita}, {Van Tent},
  {Vibert}, {Vielva}, {Villa}, {Vittorio}, {Wade}, {Wandelt}, {White}, {White},
  {Yvon}, {Zacchei}, \& {Zonca}}]{2014planck}
{Planck Collaboration}, {Ade}, P.~A.~R., {Aghanim}, N., {et~al.} 2014, \aap,
  571, A29, \dodoi{10.1051/0004-6361/201321523}

\bibitem[{{Planck Collaboration} {et~al.}(2016){Planck Collaboration}, {Ade},
  {Aghanim}, {Arnaud}, {Ashdown}, {Aumont}, {Baccigalupi}, {Banday},
  {Barreiro}, {Barrena}, {Bartlett}, {Bartolo}, {Battaner}, {Battye},
  {Benabed}, {Beno{\^\i}t}, {Benoit-L{\'e}vy}, {Bernard}, {Bersanelli},
  {Bielewicz}, {Bikmaev}, {B{\"o}hringer}, {Bonaldi}, {Bonavera}, {Bond},
  {Borrill}, {Bouchet}, {Bucher}, {Burenin}, {Burigana}, {Butler}, {Calabrese},
  {Cardoso}, {Carvalho}, {Catalano}, {Challinor}, {Chamballu}, {Chary},
  {Chiang}, {Chon}, {Christensen}, {Clements}, {Colombi}, {Colombo}, {Combet},
  {Comis}, {Couchot}, {Coulais}, {Crill}, {Curto}, {Cuttaia}, {Dahle},
  {Danese}, {Davies}, {Davis}, {de Bernardis}, {de Rosa}, {de Zotti},
  {Delabrouille}, {D{\'e}sert}, {Dickinson}, {Diego}, {Dolag}, {Dole},
  {Donzelli}, {Dor{\'e}}, {Douspis}, {Ducout}, {Dupac}, {Efstathiou},
  {Eisenhardt}, {Elsner}, {En{\ss}lin}, {Eriksen}, {Falgarone}, {Fergusson},
  {Feroz}, {Ferragamo}, {Finelli}, {Forni}, {Frailis}, {Fraisse}, {Franceschi},
  {Frejsel}, {Galeotta}, {Galli}, {Ganga}, {G{\'e}nova-Santos}, {Giard},
  {Giraud-H{\'e}raud}, {Gjerl{\o}w}, {Gonz{\'a}lez-Nuevo}, {G{\'o}rski},
  {Grainge}, {Gratton}, {Gregorio}, {Gruppuso}, {Gudmundsson}, {Hansen},
  {Hanson}, {Harrison}, {Hempel}, {Henrot-Versill{\'e}},
  {Hern{\'a}ndez-Monteagudo}, {Herranz}, {Hildebrandt}, {Hivon}, {Hobson},
  {Holmes}, {Hornstrup}, {Hovest}, {Huffenberger}, {Hurier}, {Jaffe}, {Jaffe},
  {Jin}, {Jones}, {Juvela}, {Keih{\"a}nen}, {Keskitalo}, {Khamitov}, {Kisner},
  {Kneissl}, {Knoche}, {Kunz}, {Kurki-Suonio}, {Lagache}, {Lamarre}, {Lasenby},
  {Lattanzi}, {Lawrence}, {Leonardi}, {Lesgourgues}, {Levrier}, {Liguori},
  {Lilje}, {Linden-V{\o}rnle}, {L{\'o}pez-Caniego}, {Lubin},
  {Mac{\'\i}as-P{\'e}rez}, {Maggio}, {Maino}, {Mak}, {Mandolesi}, {Mangilli},
  {Martin}, {Mart{\'\i}nez-Gonz{\'a}lez}, {Masi}, {Matarrese}, {Mazzotta},
  {McGehee}, {Mei}, {Melchiorri}, {Melin}, {Mendes}, {Mennella}, {Migliaccio},
  {Mitra}, {Miville-Desch{\^e}nes}, {Moneti}, {Montier}, {Morgante},
  {Mortlock}, {Moss}, {Munshi}, {Murphy}, {Naselsky}, {Nastasi}, {Nati},
  {Natoli}, {Netterfield}, {N{\o}rgaard-Nielsen}, {Noviello}, {Novikov},
  {Novikov}, {Olamaie}, {Oxborrow}, {Paci}, {Pagano}, {Pajot}, {Paoletti},
  {Pasian}, {Patanchon}, {Pearson}, {Perdereau}, {Perotto}, {Perrott},
  {Perrotta}, {Pettorino}, {Piacentini}, {Piat}, {Pierpaoli}, {Pietrobon},
  {Plaszczynski}, {Pointecouteau}, {Polenta}, {Pratt}, {Pr{\'e}zeau}, {Prunet},
  {Puget}, {Rachen}, {Reach}, {Rebolo}, {Reinecke}, {Remazeilles}, {Renault},
  {Renzi}, {Ristorcelli}, {Rocha}, {Rosset}, {Rossetti}, {Roudier}, {Rozo},
  {Rubi{\~n}o-Mart{\'\i}n}, {Rumsey}, {Rusholme}, {Rykoff}, {Sandri}, {Santos},
  {Saunders}, {Savelainen}, {Savini}, {Schammel}, {Scott}, {Seiffert},
  {Shellard}, {Shimwell}, {Spencer}, {Stanford}, {Stern}, {Stolyarov},
  {Stompor}, {Streblyanska}, {Sudiwala}, {Sunyaev}, {Sutton}, {Suur-Uski},
  {Sygnet}, {Tauber}, {Terenzi}, {Toffolatti}, {Tomasi}, {Tramonte},
  {Tristram}, {Tucci}, {Tuovinen}, {Umana}, {Valenziano}, {Valiviita}, {Van
  Tent}, {Vielva}, {Villa}, {Wade}, {Wandelt}, {Wehus}, {White}, {Wright},
  {Yvon}, {Zacchei}, \& {Zonca}}]{2016planck}
---. 2016, \aap, 594, A27, \dodoi{10.1051/0004-6361/201525823}

\bibitem[{{Proctor} {et~al.}(2015){Proctor}, {Mendes de Oliveira}, {Azanha},
  {Dupke}, \& {Overzier}}]{2015proctor}
{Proctor}, R.~N., {Mendes de Oliveira}, C., {Azanha}, L., {Dupke}, R., \&
  {Overzier}, R. 2015, \mnras, 449, 2345, \dodoi{10.1093/mnras/stv371}

\bibitem[{{Randall} {et~al.}(2008){Randall}, {Markevitch}, {Clowe}, {Gonzalez},
  \& {Brada{\v{c}}}}]{2008randall}
{Randall}, S.~W., {Markevitch}, M., {Clowe}, D., {Gonzalez}, A.~H., \&
  {Brada{\v{c}}}, M. 2008, \apj, 679, 1173, \dodoi{10.1086/587859}

\bibitem[{{Randall} {et~al.}(2016){Randall}, {Clarke}, {van Weeren}, {Intema},
  {Dawson}, {Mroczkowski}, {Blanton}, {Bulbul}, \& {Giacintucci}}]{2016randall}
{Randall}, S.~W., {Clarke}, T.~E., {van Weeren}, R.~J., {et~al.} 2016, \apj,
  823, 94, \dodoi{10.3847/0004-637X/823/2/94}

\bibitem[{{Schlafly} \& {Finkbeiner}(2011)}]{2011schlafly}
{Schlafly}, E.~F., \& {Finkbeiner}, D.~P. 2011, \apj, 737, 103,
  \dodoi{10.1088/0004-637X/737/2/103}

\bibitem[{{Schrabback} {et~al.}(2018){Schrabback}, {Applegate}, {Dietrich},
  {Hoekstra}, {Bocquet}, {Gonzalez}, {von der Linden}, {McDonald}, {Morrison},
  {Raihan}, {Allen}, {Bayliss}, {Benson}, {Bleem}, {Chiu}, {Desai}, {Foley},
  {de Haan}, {High}, {Hilbert}, {Mantz}, {Massey}, {Mohr}, {Reichardt}, {Saro},
  {Simon}, {Stern}, {Stubbs}, \& {Zenteno}}]{2018schrabback}
{Schrabback}, T., {Applegate}, D., {Dietrich}, J.~P., {et~al.} 2018, \mnras,
  474, 2635, \dodoi{10.1093/mnras/stx2666}

\bibitem[{{Seitz} \& {Schneider}(1997)}]{1997seitz}
{Seitz}, C., \& {Schneider}, P. 1997, \aap, 318, 687.
\newblock \doarXiv{astro-ph/9601079}

\bibitem[{{Shupe} {et~al.}(2012){Shupe}, {Laher}, {Storrie-Lombardi}, {Surace},
  {Grillmair}, {Levitan}, \& {Sesar}}]{2012shupe}
{Shupe}, D.~L., {Laher}, R.~R., {Storrie-Lombardi}, L., {et~al.} 2012, in
  Society of Photo-Optical Instrumentation Engineers (SPIE) Conference Series,
  Vol. 8451, Software and Cyberinfrastructure for Astronomy II, ed. N.~M.
  {Radziwill} \& G.~{Chiozzi}, 84511M, \dodoi{10.1117/12.925460}

\bibitem[{{Sirianni} {et~al.}(2005){Sirianni}, {Jee}, {Ben{\'\i}tez},
  {Blakeslee}, {Martel}, {Meurer}, {Clampin}, {De Marchi}, {Ford}, {Gilliland
  }, {Hartig}, {Illingworth}, {Mack}, \& {McCann}}]{2005sirianni}
{Sirianni}, M., {Jee}, M.~J., {Ben{\'\i}tez}, N., {et~al.} 2005, \pasp, 117,
  1049, \dodoi{10.1086/444553}

\bibitem[{Umetsu {et~al.}(2011)Umetsu, Broadhurst, Zitrin, Medezinski, Coe, \&
  Postman}]{2011umetsu}
Umetsu, K., Broadhurst, T., Zitrin, A., {et~al.} 2011, The Astrophysical
  Journal, 738, 41, \dodoi{10.1088/0004-637x/738/1/41}

\bibitem[{{van Weeren} {et~al.}(2019){van Weeren}, {de Gasperin}, {Akamatsu},
  {Br{\"u}ggen}, {Feretti}, {Kang}, {Stroe}, \& {Zandanel}}]{2019vanweeren}
{van Weeren}, R.~J., {de Gasperin}, F., {Akamatsu}, H., {et~al.} 2019, \ssr,
  215, 16, \dodoi{10.1007/s11214-019-0584-z}

\bibitem[{{Vikhlinin} {et~al.}(2009){Vikhlinin}, {Kravtsov}, {Burenin},
  {Ebeling}, {Forman}, {Hornstrup}, {Jones}, {Murray}, {Nagai}, {Quintana}, \&
  {Voevodkin}}]{2009vikhlinin}
{Vikhlinin}, A., {Kravtsov}, A.~V., {Burenin}, R.~A., {et~al.} 2009, \apj, 692,
  1060, \dodoi{10.1088/0004-637X/692/2/1060}

\bibitem[{{Wright} \& {Brainerd}(2000)}]{2000wright}
{Wright}, C.~O., \& {Brainerd}, T.~G. 2000, \apj, 534, 34,
  \dodoi{10.1086/308744}

\bibitem[{{Zitrin} {et~al.}(2012){Zitrin}, {Bartelmann}, {Umetsu}, {Oguri}, \&
  {Broadhurst}}]{2012zitrin}
{Zitrin}, A., {Bartelmann}, M., {Umetsu}, K., {Oguri}, M., \& {Broadhurst}, T.
  2012, \mnras, 426, 2944, \dodoi{10.1111/j.1365-2966.2012.21886.x}

\end{thebibliography}
\bibliographystyle{aasjournal}


%% Include this line if you are using the \edit1, \edit1, \deleted
\listofchanges

\end{document}